\documentclass[]{aa}  

\usepackage{graphicx}
\usepackage[varg]{txfonts}
\usepackage{amsmath}
\usepackage{natbib}
\usepackage[modulo,switch]{lineno}

\newcommand{\derivp} [2] {\frac {\partial #1 } {\partial #2} }
\newcommand{\deriv} [2] {\frac {\textrm{d} #1 } {\textrm{d} #2} }

\newcommand{\eq}[1] {Eq.\,(\ref{#1})}
\newcommand{\eqn} [1] {
\begin{equation} #1
\end{equation}}
\newcommand{\eqna} [1] {
\begin{eqnarray} #1
\end{eqnarray}}
\def\pmb#1{\mbox{\boldmath$#1$}}
\def\gtsim {>\kern-1.2em\lower1.1ex\hbox{$\sim$}}
\def\ltsim {<\kern-1.2em\lower1.1ex\hbox{$\sim$}}
\def\gtsim {>\kern-1.2em\lower1.1ex\hbox{$\sim$}}
\def\ltsim {<\kern-1.2em\lower1.1ex\hbox{$\sim$}}

\def\be{\begin{equation}}
\def\ee{\end{equation}}

\begin{document}

\title{Transport of angular momentum by stochastically excited waves as an explanation for the outburst of the rapidly rotating Be star HD49330\thanks{Based on observations obtained with the CoRoT satellite.}}
\titlerunning{Transport of angular momentum in the outbursting Be star HD\,49330}

\author{C. Neiner\inst{1}
\and U. Lee\inst{2}
\and S. Mathis\inst{3,1}
\and H. Saio\inst{2}
\and C. C. Lovekin\inst{4}
\and K. C. Augustson\inst{3}
}

\offprints{C. Neiner}

\institute{
LESIA, Paris Observatory, PSL University, CNRS, Sorbonne
Universit\'e, Universit\'e de Paris, 5 place Jules Janssen, 92195 Meudon, France,
\email{coralie.neiner@obspm.fr}
\and Astronomical Institute, Graduate School of Science, Tohoku University, Sendai, 980-8578, Japan
\and AIM, CEA, CNRS, Université Paris-Saclay, Université Paris Diderot, Sorbonne Paris Cité, F-91191 Gif-sur-Yvette Cedex, France
\and Physics Department, Mount Allison University, Sackville, NB, E4L 1E6, Canada
}

\date{Received ...; accepted ...}
 
\abstract
{HD\,49330 is an early Be star that underwent an outburst during its five-month observation with the CoRoT satellite. An analysis of its light curve revealed several independent p and g pulsation modes, in addition to showing that the amplitude of the modes is directly correlated with the outburst.}
{We modelled the results obtained with CoRoT to understand the link between pulsational parameters and the outburst of this Be star.}
{We modelled the flattening of the structure of the star due to rapid rotation in two ways: Chandrasekhar-Milne's expansion and 2D structure computed with {{\sc{ROTORC}}}. We then modelled $\kappa$-driven pulsations. We also adapted the formalism of the excitation and amplitude of stochastically excited gravito-inertial modes to rapidly rotating stars, and we modelled those pulsations as well.}
{We find that while pulsation p modes are indeed excited by the $\kappa$ mechanism, the observed g modes are, rather, a result of stochastic excitation. In contrast, g and r waves are stochastically excited in the convective core and transport angular momentum to the surface, increasing its rotation rate. This destabilises the external layers of the star, which then emits transient stochastically excited g waves. These transient waves produce most of the low-frequency signal detected in the CoRoT data and ignite the outburst. During this unstable phase, p modes disappear at the surface because their cavity is broken. Following the outburst and ejection of the surface layer, relaxation occurs, making the transient g waves disappear and p modes reappear.}
{This work includes the first coherent model of stochastically excited gravito-inertial pulsation modes in a rapidly rotating Be star. It provides an explanation for the correlation between the variation in the amplitude of frequencies detected in the CoRoT data and the occurrence of an outburst. This scenario could apply to other pulsating Be stars, providing an explanation to the long-standing questions surrounding Be outbursts and disks.}

\maketitle
%

\section{Introduction}
\label{intro}

Be stars are late-O to early-A stars surrounded by a decretion disk fed by discrete mass-loss events \citep[see the review by][]{rivinius2013}. The ejection of material from the star into the disk requires enough angular momentum to reach the breakup velocity at the stellar surface. Most of the angular momentum is provided by the rapid rotation of the star \citep[on average, 90\% of the critical velocity in our galaxy, see][]{fremat}. The physical process leading to the transport of additional angular momentum to the surface has yet to be identified. However, transport by pulsation modes has been put forwards as the most probable explanation up to now \citep{lee_saio,pantillon2007,mathis}. Indeed, Be stars are known to host pulsations excited by the $\kappa$ mechanism, similar to those observed in $\beta$\,Cephei and SPB stars \citep[e.g.][]{neiner2009}. More recently, pulsations that are excited stochastically have also been observed in Be stars \citep{neinerStocha} and in O stars \citep{aerts2015}. Gravity waves are also found to be excited in numerical simulations, to efficiently transport angular momentum in hot stars \citep{rogers2013}, and to provide predictions for rotation profiles that match the seismic observations \citep{rogers2015}. Moreover, theoretical calculations have shown that rotation is a crucial ingredient for exciting gravito-inertial waves \citep{MNT2014}.

HD\,49330 is a rapidly rotating B0.5IVe star observed with the CoRoT satellite
\citep{auvergne} from October 18, 2007 to March 3, 2008, that is over  $\sim$137 consecutive days with a gap of 3.5 days. During the CoRoT observation, an
outburst of several hundredths of magnitude occurred. Outbursts, and their subsequent dimming, are 
known to occur frequently in this star. This was, however, the first time that an
outburst could be observed with an extremely high photometric accuracy and
cadence. 

The analysis of the CoRoT light curve of HD\,49330 \citep{huat} and the
associated ground-based spectroscopic data \citep{floquet} led to the detection
of 30 independent groups of frequencies of variation composed of more than 300
frequencies in total. These frequencies were interpreted in terms of p and g
pulsation modes. The amplitude of the various modes are directly correlated with
the phases of the outburst: p modes have a higher amplitude during the quiescent
and relaxation phases, while g modes increase in amplitude and become prominent 
just before (precursor phase)
and during the outburst; see \cite{huat} for more details. Although the
correlation between the changes in the pulsation parameters and the outburst is
clear, the mechanism linking the two phenomena still needs to be investigated.
This can be done thanks to seismic modelling and this investigation is the purpose of this paper.

To create a seismic model, we need to know the pulsational parameters as well
as the fundamental parameters of the star. The pulsational parameters are
provided in \cite{huat}. \cite{floquet} (see their Table 2) determined that for
$\frac{\Omega}{\Omega_c}$=0.9 at the equator, $T_{\rm eff}$=28000 $\pm$ 1500 K, $\log g$=3.9
$\pm$ 0.2, $v{\sin}i$=280 $\pm$ 40 km.s$^{-1}$, $i$=43 $\pm$ 11 deg, $M$=14.4
$\pm$ 0.3 $M_{\odot}$, $R_{\rm eq}$ = 8.7 $\pm$ 2.5 $R_{\odot}$, and $\log L$=4.5
$\pm$ 0.2 $L_{\odot}$. While $\frac{\Omega}{\Omega_c}$ is commonly used in observational studies, seismic models rather consider $\overline\Omega$, which is the rotation frequency normalised by $\sqrt{GM/R^3}$, where R is the mean stellar radius. For HD\,49330,
$\frac{\Omega}{\Omega_c}$=0.9 corresponds to $\overline\Omega$=0.7833.

In Sect.~\ref{sect_kappa}, we explain the models we use for the structure of
the rapidly rotating star and present our models of $\kappa$-driven pulsation modes. We then
propose in Sect.~\ref{stochastic} an alternative explanation for the CoRoT
observations based on stochastically-driven g waves. We discuss the excitation of such
waves and present seismic modelling results, including stochastic excitation. We  discuss the angular momentum transport by those waves and the ignition of the outburst.
Finally, we draw conclusions about the treatment and interpretation of seismic
observations of early Be stars and about the impact of stochastic g waves on the
ignition of Be outbursts in Sect.~\ref{conclusion}.

\section{Seismic modelling of HD\,49330 with the classical $\kappa$ mechanism}
\label{sect_kappa}

The modelling of the pulsations of a star first requires a calculation of the
structure of the star and then a pulsation stability analysis.

\subsection{Modelling the rapidly rotating stellar structure}
\label{sect_structure}

\begin{figure}[!ht]
\begin{center}
\resizebox{\hsize}{!}{\includegraphics[trim={0.5cm 4.2cm 0.3cm 4.3cm},clip]{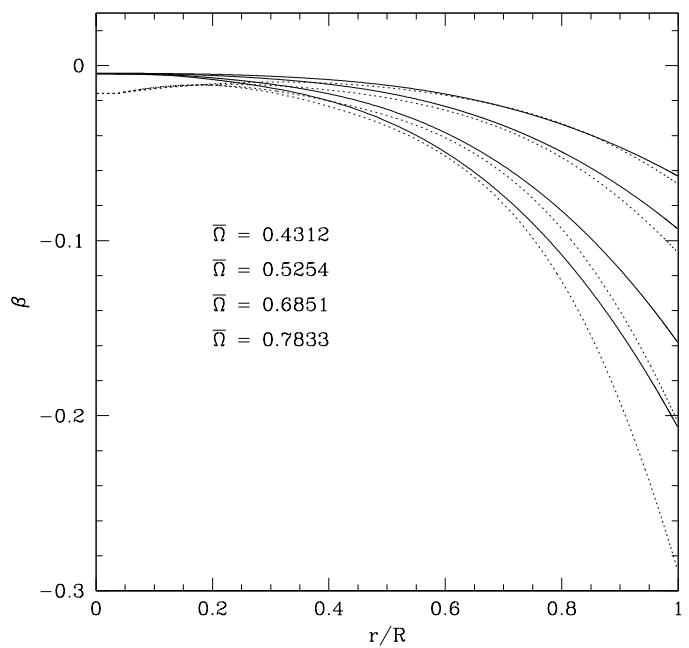}}
\caption[]{Coefficient $\beta$ of the Legendre function $P_2(\cos\theta)$ as a function
of the fractional radius for a 10$M_\odot$ star rotating at various rates. Solid
lines represent the deformation from the Chandrasekhar-Milne's expansion and
dotted lines the deformation from 2D structure models. $\overline\Omega$ is the rotation frequency normalised by $\sqrt{GM/R^3}$ where R is the mean radius.
$\overline\Omega$ increases from top to bottom in the figure. $\overline\Omega$=0.7833 corresponds to $\frac{\Omega}{\Omega_c}$=0.9 at the equator as observed for HD\,49330.}
\label{deformation}
\end{center}
\end{figure}

For the stellar structure we use two different approaches: 
\begin{enumerate}
\item A 2D structure model calculated with the {\sc ROTORC} code \citep{deupree90, deupree95}. {\sc ROTORC} solves the conservation equations for mass, momentum, energy and composition, as well as Poisson’s equation on a 2D spherical grid. The model rotates uniformly on the zero age main sequence and the interior angular momentum is conserved locally as the model evolves. The surface is assumed to be an equipotential. The independent variables are the fractional surface equatorial radius and the colatitude. This allows us to obtain the rotationally flattened structure of the star.
\item A deformation following Chandrasekhar-Milne's expansion, in which the effects of rotation on the shape of the star are taken into account with a method similar to that of \cite{lee_baraffe} but slightly modified with the 2D model as input. In this case an isobaric surface is expressed as $r = a[1 + \alpha(a) + \beta(a)P_2(\cos\theta)]$ where $a$ is the mean radius, $P_2(\cos\theta)$ is the Legendre function, and the functions $\alpha$ and $\beta$ represent the horizontal averaged effects of the centrifugal force and the deformation of the equilibrium state, respectively. Here we obtain the value of $\beta(a)$ for a given rotation frequency by interpolating the coefficients determined from the 2D {\sc ROTORC} model. This modification is necessary because Chandrasekhar-Milne's expansion is only valid when the difference between the rotating and non-rotating structures can be approximated by $\Omega^2$ order corrections, which is not the case when the rotation velocity is very high. 
\end{enumerate}
Details about the above two approaches are described in Sect.~3 of \cite{neinerMixing}.

Figure~\ref{deformation} shows the coefficient $\beta$ as a function of the fractional radius for a 10$M_\odot$ star rotating at various rates. The figure shows that for relatively slow rotation rates, the Chandrasekhar-Milne expansion and the 2D structure model are similar. For $\overline\Omega$ $>$ 0.5, however, the deformations from the 2D models are considerably larger than those from Chandrasekhar-Milne's expansion, in particular in the outer layers.  

\subsection{Tohoku models of $\kappa$-driven pulsations}

To model the pulsations, we use the Tohoku non-adiabatic pulsation code. This code includes the effects of the centrifugal and of the Coriolis accelerations. With this code, we perform a non-radial pulsation stability analysis for HD\,49330 models. This analysis allows us to predict the modes that will be excited by the $\kappa$-mechanism as well as their frequencies and growth rates. More details of this oscillation code are described in \cite{neinerMixing}. 

\begin{figure*}[!ht]
\begin{center}
\resizebox{0.48\hsize}{!}{\includegraphics[clip]{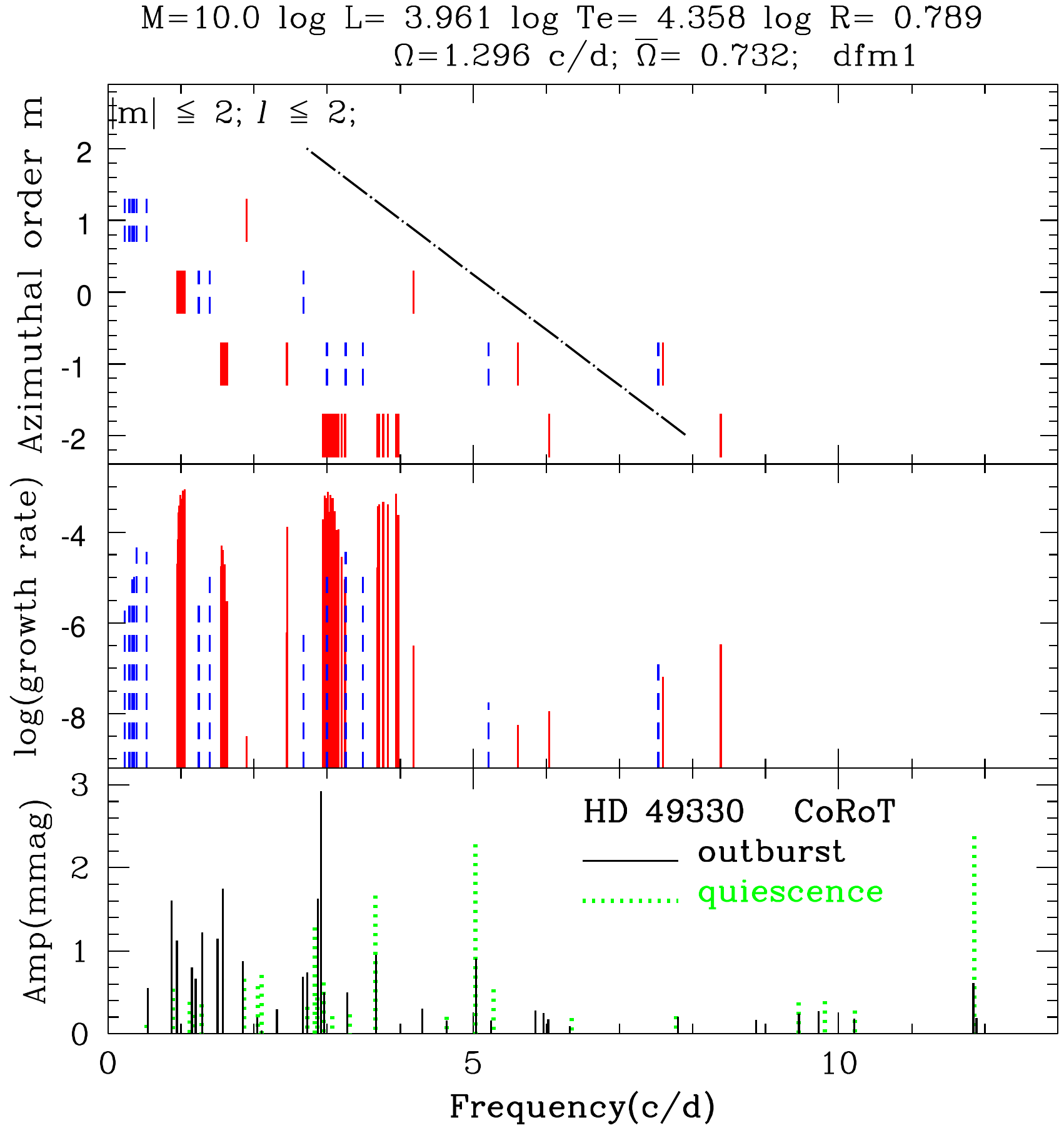}}
\resizebox{0.48\hsize}{!}{\includegraphics[clip]{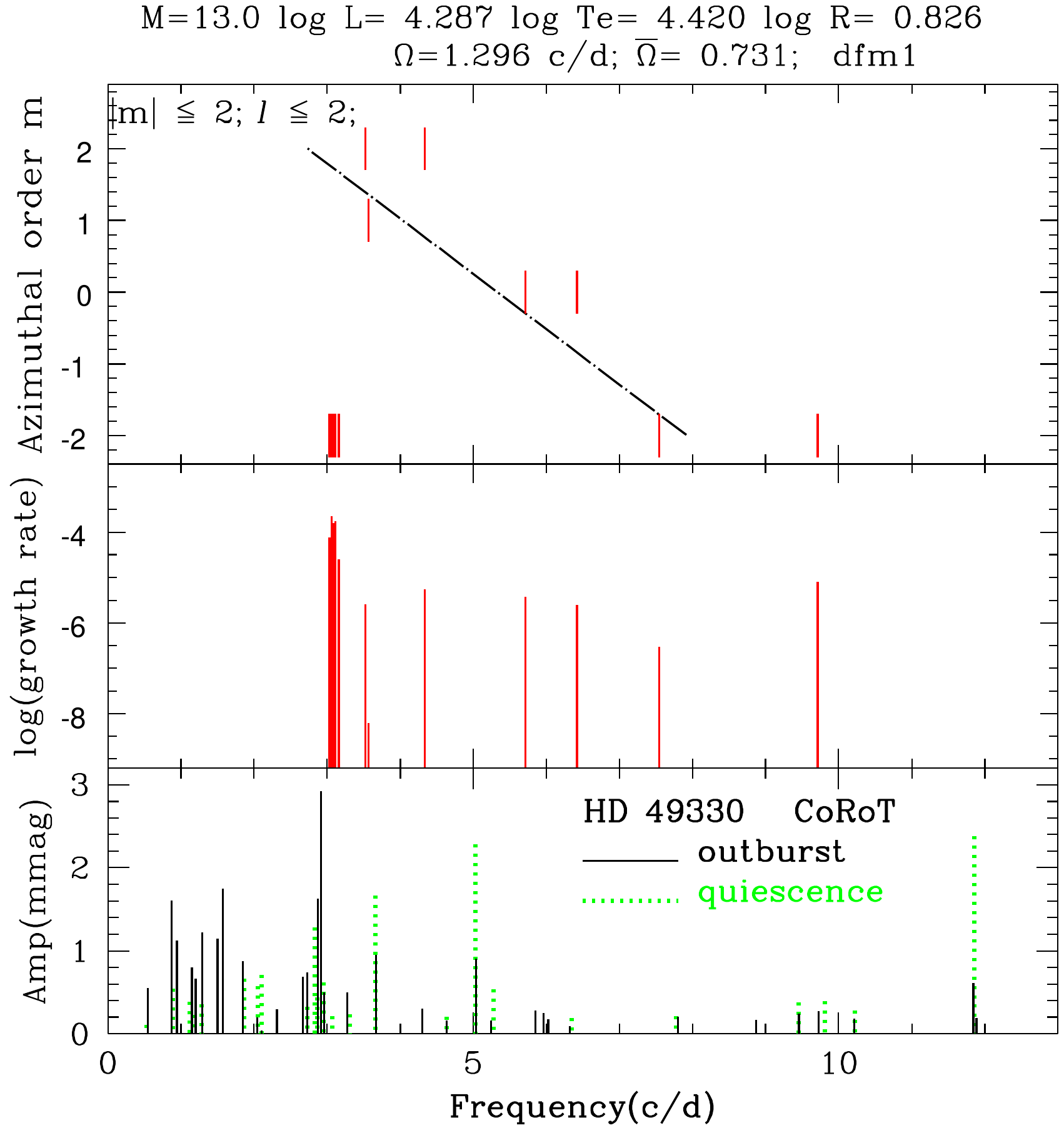}}
\caption[]{Statistically significant observed frequencies of HD\,49330 (bottom panel) during and outside
of the outburst phase compared to the modelled frequencies in a nearly
critically rotating B star (middle panel) of 10 $M_\odot$ (left) and 13 $M_\odot$ (right) calculated using rotational deformation based on Chandrasekhar-Milne's expansion. The top panel shows the azimuthal order $m$ of each frequency peak. The dashed line represents the approximate separation between p and g modes. Red full lines correspond to even modes and dashed blue lines to odd modes.}
\label{model_Chandra}
\end{center}
\end{figure*}

We first perform a non-adiabatic
pulsation analysis for rapidly rotating star models whose evolutionary track
pass close to the position of HD\,49330 in the HR diagram. The evolutionary
models are computed in the same manner as in \cite{walker} and the stability
analysis is based on the method by \cite{lee_baraffe}. We use the two stellar
structures presented above as an input to the models. 

We adopt the convention that a negative or positive $m$ order represents a
prograde or retrograde mode, respectively. 
Modes with even or odd parity, which corresponds to even or odd values of $l + |m|$, 
respectively possess either a symmetry or an antisymmetry about the equator of the star. 

In a rapidly rotating star, a mode cannot be represented by a single $l$ value
for a given $m$ order because of the latitudinal couplings between the various
spherical harmonics owing to the Coriolis and the centrifugal accelerations. 
Thus, each mode must be expanded over as a series of spherical harmonics.  Consequently, for each mode we choose the component of this series with the highest amplitude at the stellar surface and use its $l$ value to compare with the $l$ value determined from the observations.

It is not our goal with these models to reproduce each individual observed
frequency in HD\,49330 but, rather, to reproduce the ranges of frequencies excited in the
star.

We compute non-adiabatic models using a stellar structure deformation based
either on Chandrasekhar-Milne's expansion or on a 2D stellar structure, as
presented in Sect.~\ref{sect_structure}.

\begin{figure*}[!ht]
\begin{center}
\resizebox{0.48\hsize}{!}{\includegraphics[clip]{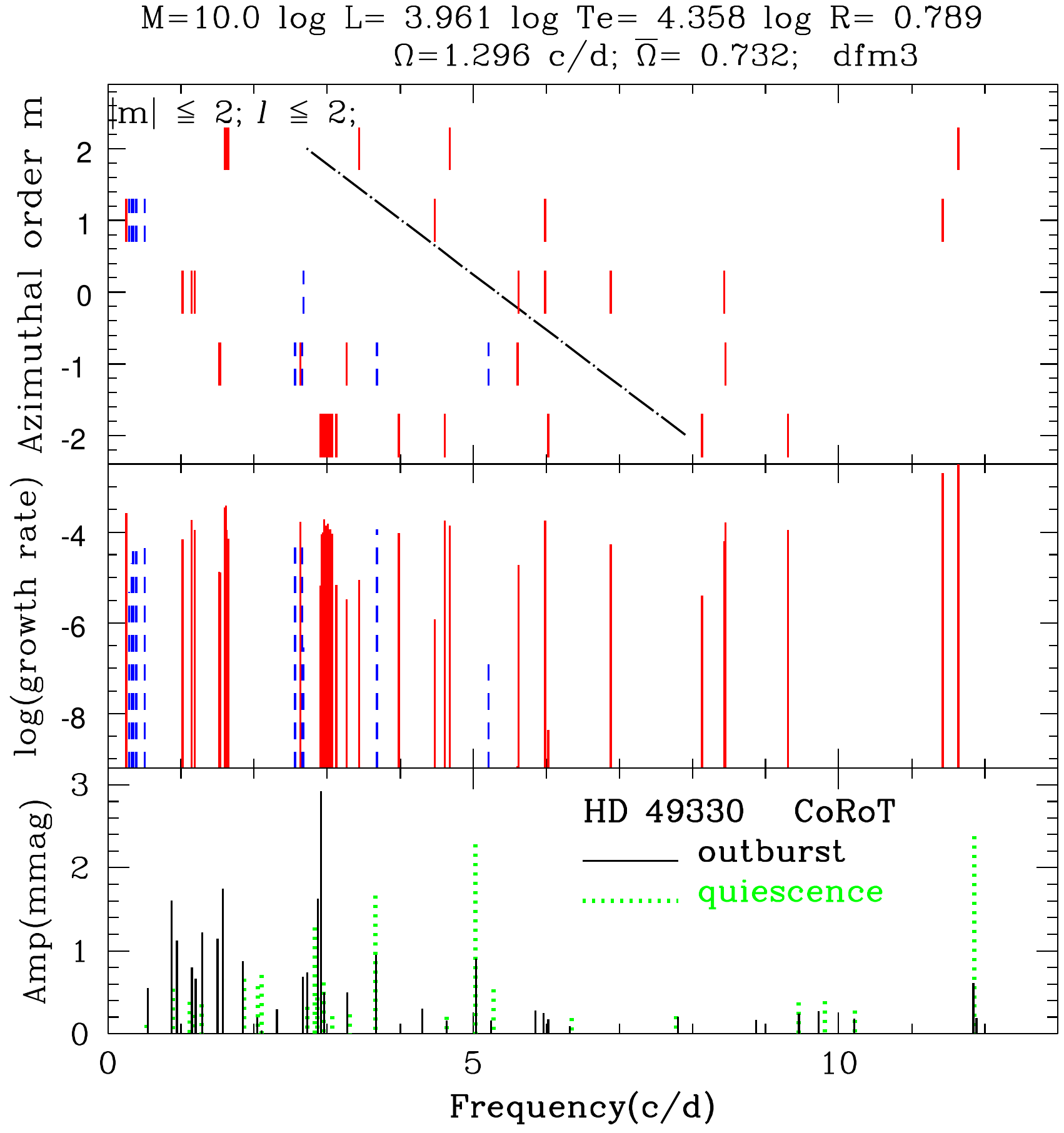}}
\resizebox{0.48\hsize}{!}{\includegraphics[clip]{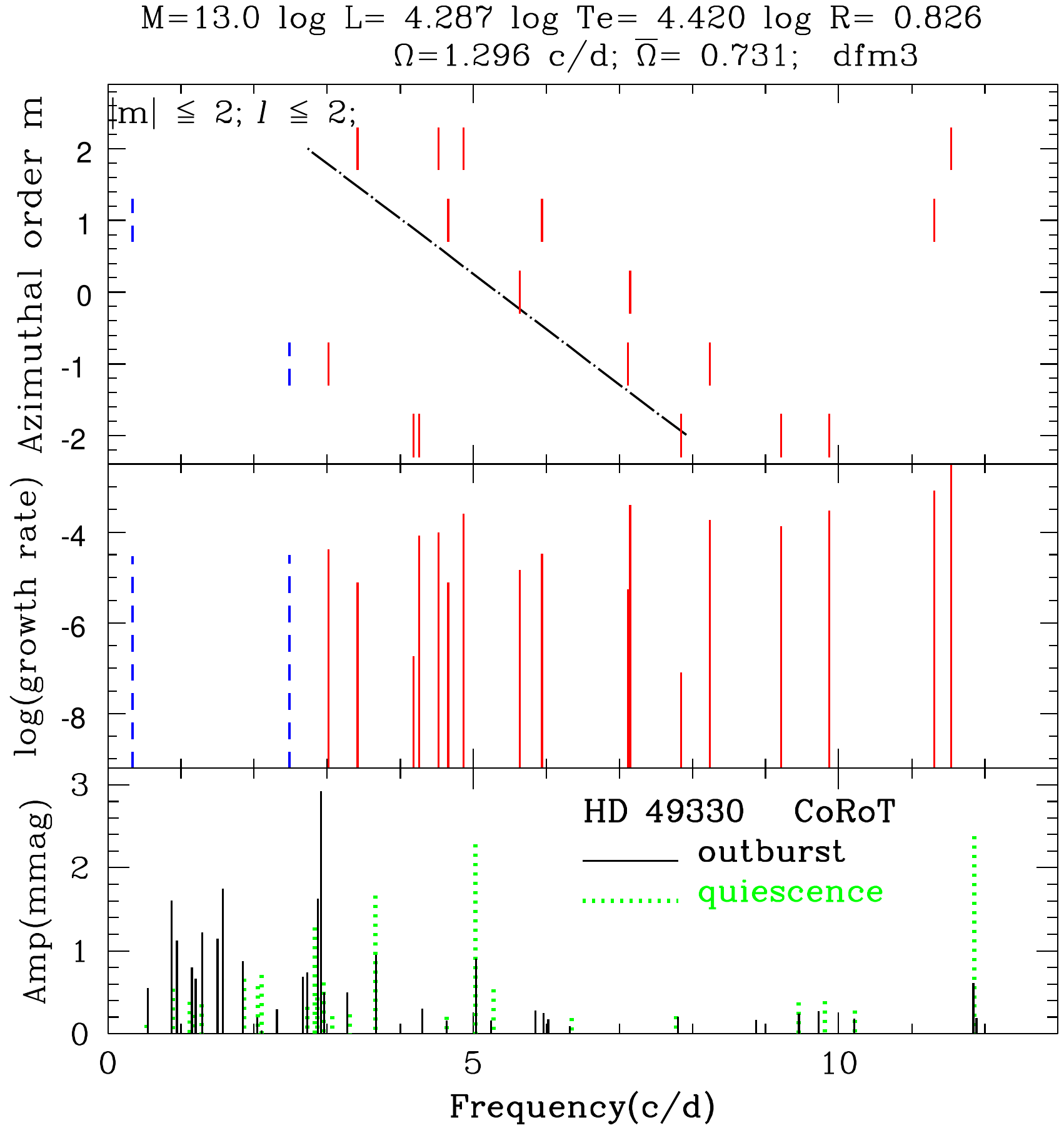}}
\caption[]{Same as Fig.~\ref{model_Chandra} but the modelled frequencies are
calculated using a 2D structure model.}
\label{model_2D}
\end{center}
\end{figure*}

Using Chandrasekhar-Milnes's expansion, our models reproduce g modes in the observed
frequency range only for models with masses significantly lower than the mass determined from spectroscopic observations. See examples in Fig.~\ref{model_Chandra} for a 10 and 13 $M_\odot$ star: the bottom panel shows the frequencies observed in the CoRoT data during or outside of the outburst detected by \cite{huat} and which have a signal-to-noise ratio (S/N) above 4, the middle panel shows the growth rate of the modelled modes, the top panel shows the azimuthal order $m$ with colors indicating the even and odd parity modes. The typical growth rates of excited modes are 10$^{-4}$, meaning that it would take 10$^4$ times the oscillation period for the amplitude to grow. g modes are excited in the observed frequency range for the lower mass model but not for M = 13 $M_\odot$. On the contrary, p modes are excited for the higher mass model but not at lower masses. It is thus difficult to excite both p and g modes at the same time whatever the set of stellar parameters with the Chandrasekhar-Milnes's expansion.

Using rotational deformation based on 2D structure models, both p and g modes
can be excited at the same time. Indeed, g modes seem rather insensitive to the
way we treat stellar rotational deformation. The excitation of high-frequency
p modes, however, seems to occur only when the rotation is nearly critical and
thus the treatment of the rotational deformation is very important. See the left panel of Fig.~\ref{model_2D} for an example of a star of 10 $M_\odot$ with both p and g modes excited, which is to be compared with Fig.~\ref{model_Chandra}. In this
model, f=11.9 c~d$^{-1}$ can be identified with a retrograde p mode with $m$=1
or 2, while the frequency groups around 1.5 and 3 c~d$^{-1}$ are identified with
prograde g modes of $m$=-1 and -2 respectively. However, the stellar parameters
used here do not correspond to the ones derived from spectroscopy
\citep{floquet}.

With stellar parameters compatible with the error boxes derived from
spectroscopic observations (see Sect.~\ref{intro}), and with rotational deformation
extracted from 2D structure models, p modes are easily excited but g modes are
less easy to excite. The right panel of Fig.~\ref{model_2D} shows that  already for a 13 $M_\odot$ (and higher), the group of observed g modes around 1.5 c.d$^{-1}$ is not present anymore in the model. In addition, the model predicts several strong modes between 6 and 10 c.d$^{-1}$ which are not observed in the CoRoT data.

\begin{figure}[!ht]
\begin{center}
\resizebox{\hsize}{!}{\includegraphics[clip]{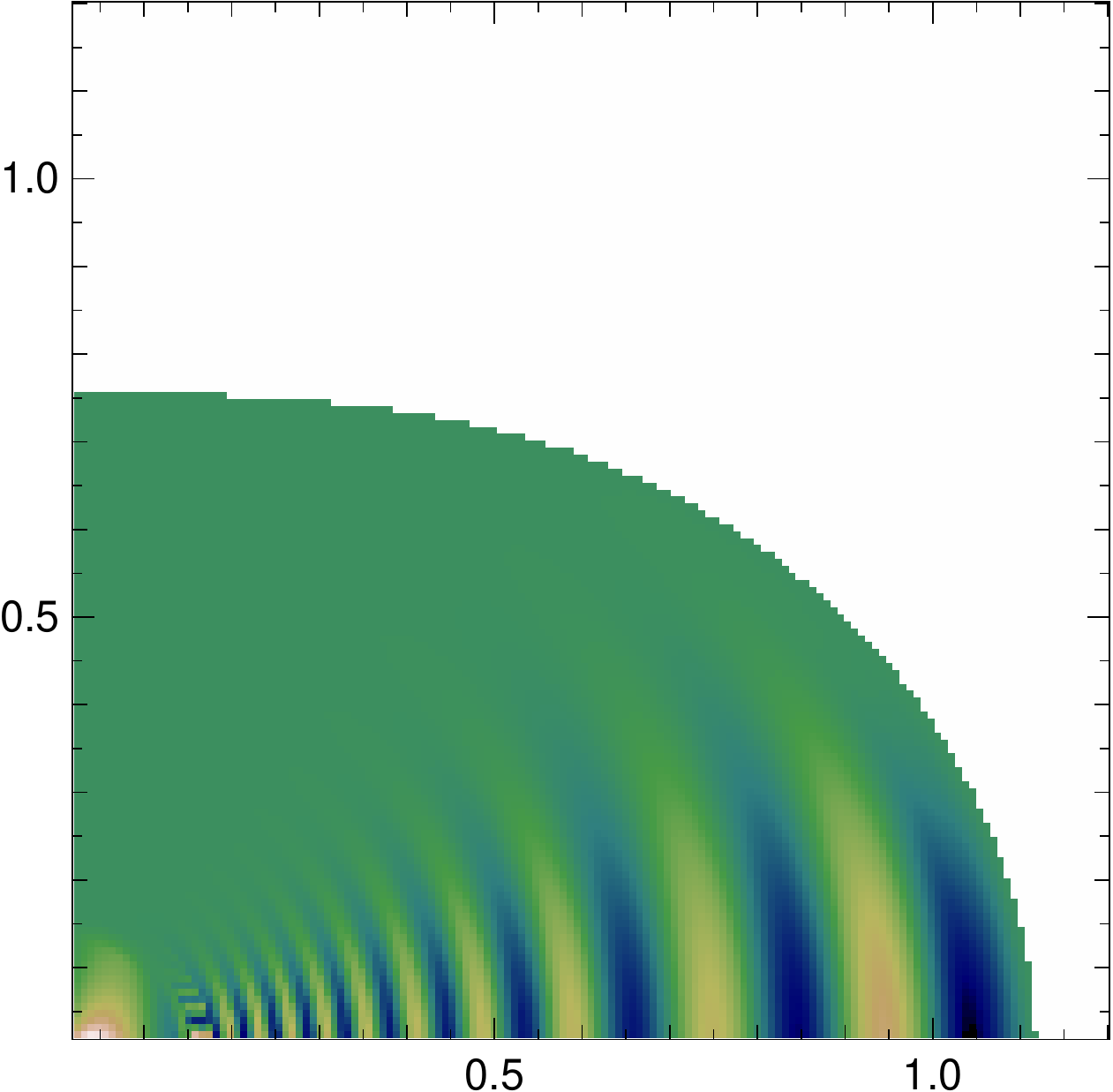}\includegraphics[clip]{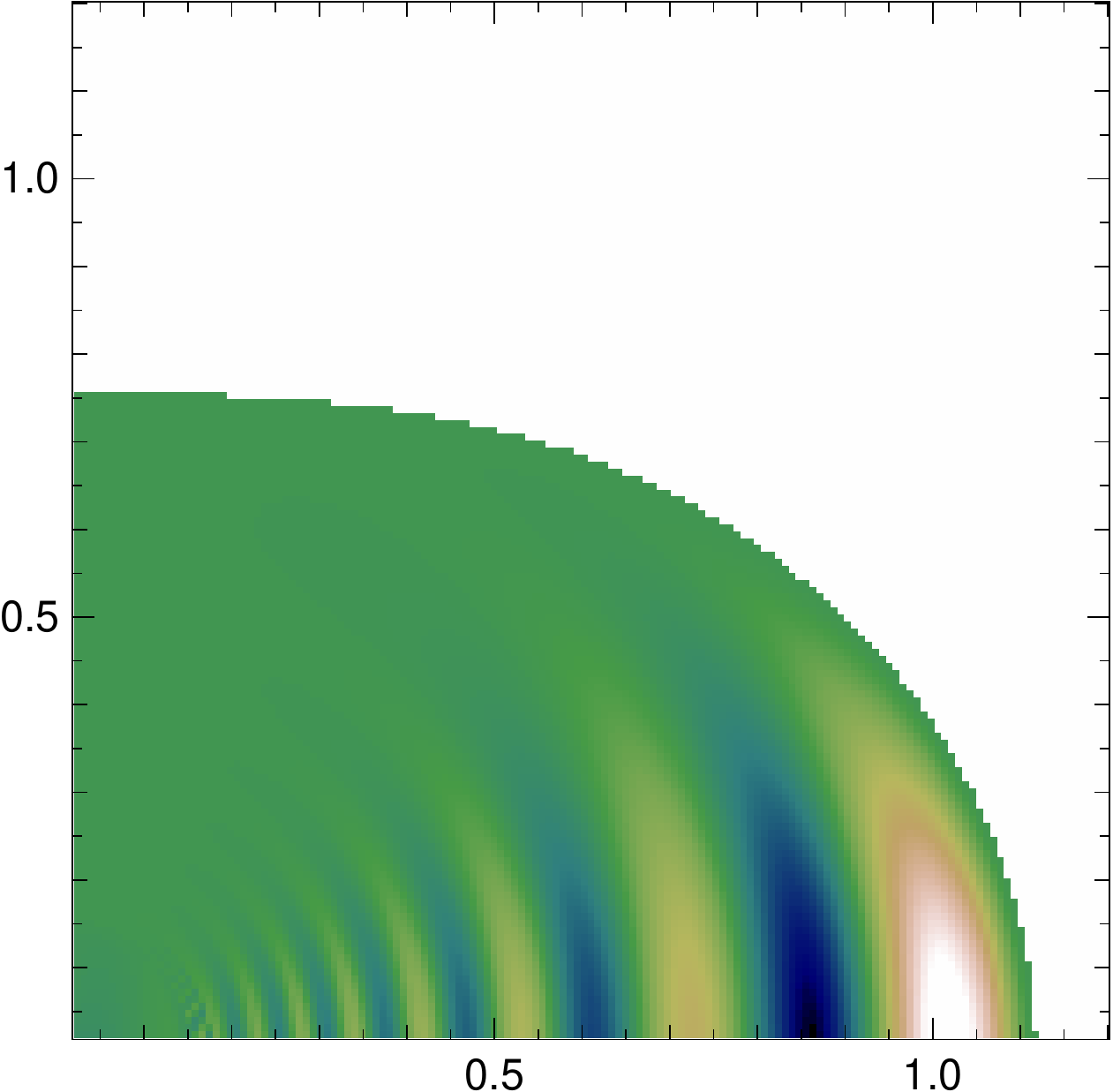}}
\caption[]{Colour maps of pulsation amplitudes for the g modes with $l=2$, $m=-2$
and $f$=2.9 $c.d^{-1}$ (left) and $l=1$, $m=-1$ and $f$=1.5 $c.d^{-1}$ (right),
the two main frequencies observed in the CoRoT data during the outburst of
HD\,49330 \citep{huat}.}
\label{map_g}
\end{center}
\end{figure}

\begin{figure}[!ht]
\begin{center}
\resizebox{\hsize}{!}{\includegraphics[clip]{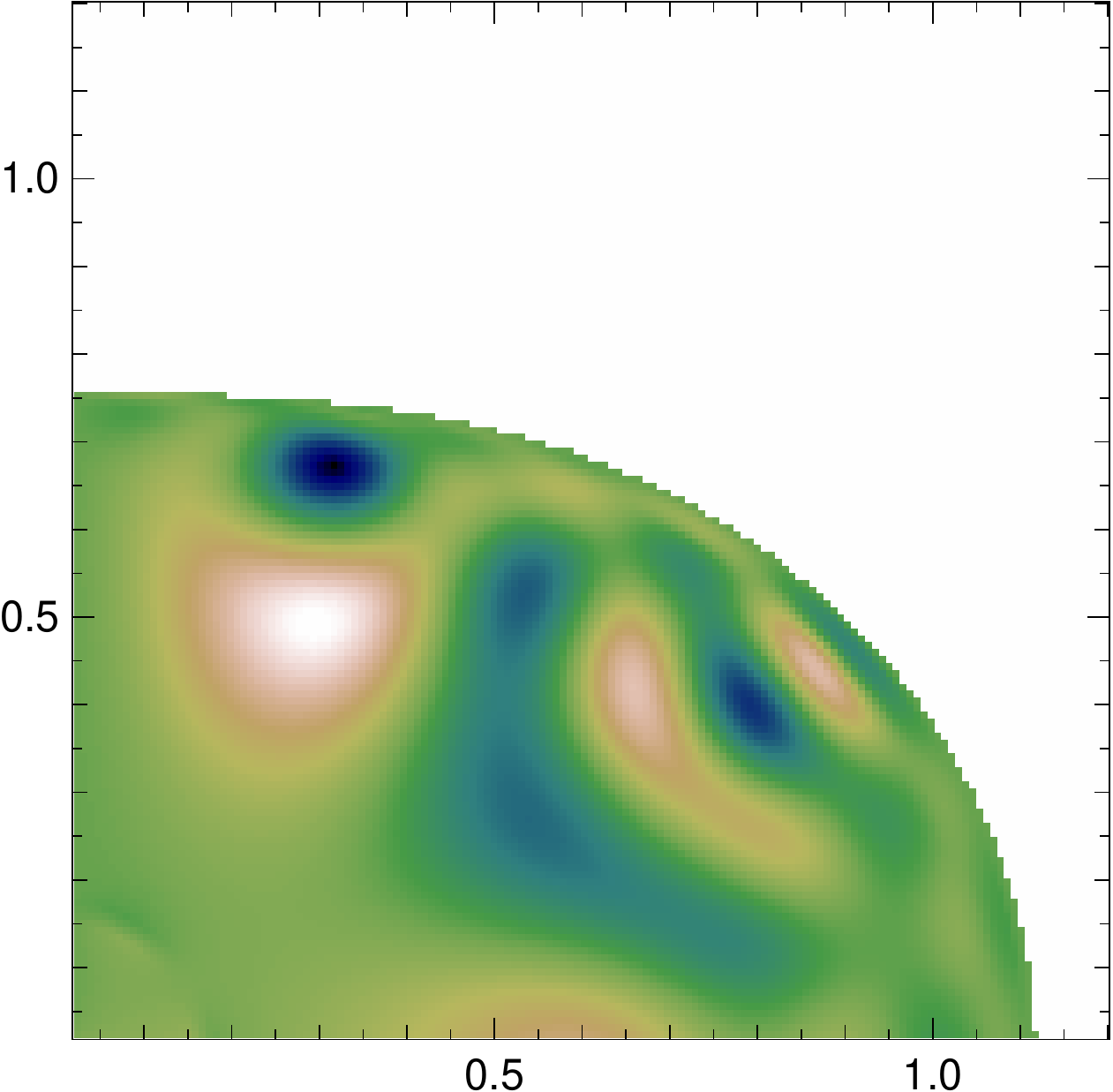}\includegraphics[clip]{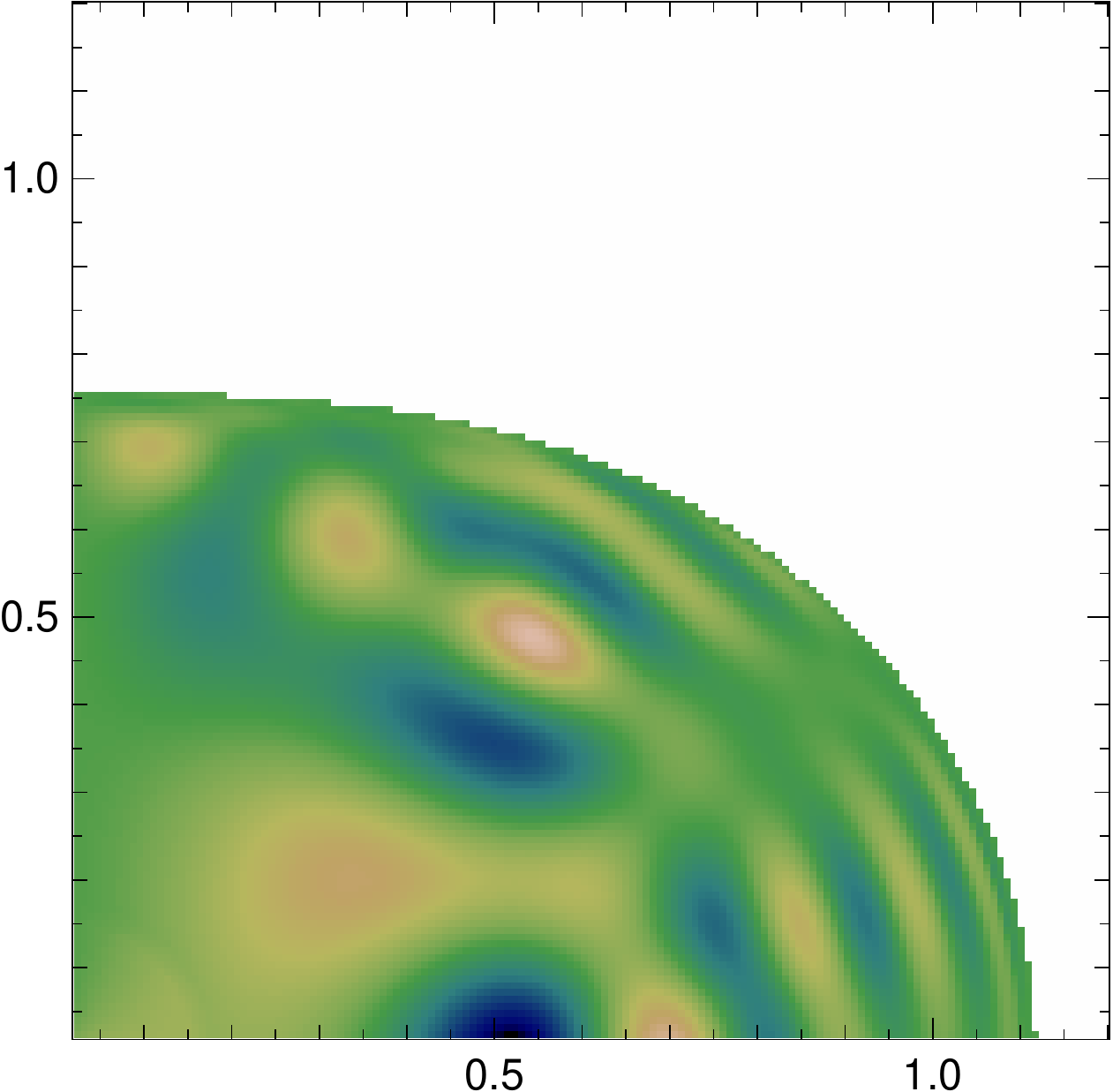}}
\caption[]{Colour maps of pulsation amplitudes for the p modes with $l=2$, $m=2$
and $f$=11.7 $c.d^{-1}$ (left) and $l=2$, $m=2$ and $f$=16.9 $c.d^{-1}$ (right),
the two frequencies observed both in the CoRoT data in the quiet phase of
HD\,49330 \citep{huat} and in spectroscopic data \citep{floquet}.}
\label{map_p}
\end{center}
\end{figure}

Using our modelling results, we can plot colour maps of pulsation amplitudes of
modes observed in HD\,49330, assuming that they are driven by the $\kappa$ mechanism.
Plotted colour quantities are the dimensionless pressure fluctuations normalised
by gravity ($\frac{p'}{\sqrt{\rho}rg}$). 
The y and x axes respectively correspond to the distance from the centre of the star along 
the rotation axis and perpendicular to it that have been normalised by the mean radius of the star.

We find that g modes are trapped at the equator of the star (see 
Fig.~\ref{map_g}) while p modes are more evenly distributed (see
Fig.~\ref{map_p}). The trapping of $\kappa$-driven g modes is due to rapid
rotation.

\subsection{Issues with the $\kappa$ mechanism}

From the models presented above we conclude that 
(1) non-adiabatic treatment is needed, in particular to study the modes below
2$\Omega$ and to know whether the modes are excited; and (2) 
a 2D structure model is required to be able to reproduce p modes for
which the rotational deformation has a strong impact. Nevertheless, 
whatever model of stellar structure and pulsations we use, we cannot excite p and g modes at the same time
with the spectroscopically-derived stellar parameters of HD\,49330 if we
consider only pulsation modes excited by the $\kappa$ mechanism.

\cite{huat} and \cite{floquet} showed that p modes are visible in the CoRoT data
of HD\,49330 when the star is in a quiet or relaxation phase, in other words outside the
observed outburst. This is when the material at the surface of the star is close
to but not at its critical velocity. During the precursor and outburst phase, g
modes appear in the CoRoT data. This is when the velocity of the material at the surface of the
star at its equator becomes critical and material gets ejected from the star.

Therefore several possibilities can be considered to try to answer the problem of excitation of both p and g modes:

\begin{figure*}[!ht]
\begin{center}
\resizebox{\hsize}{!}{\includegraphics[trim=0 2cm 4.5cm 0, clip]{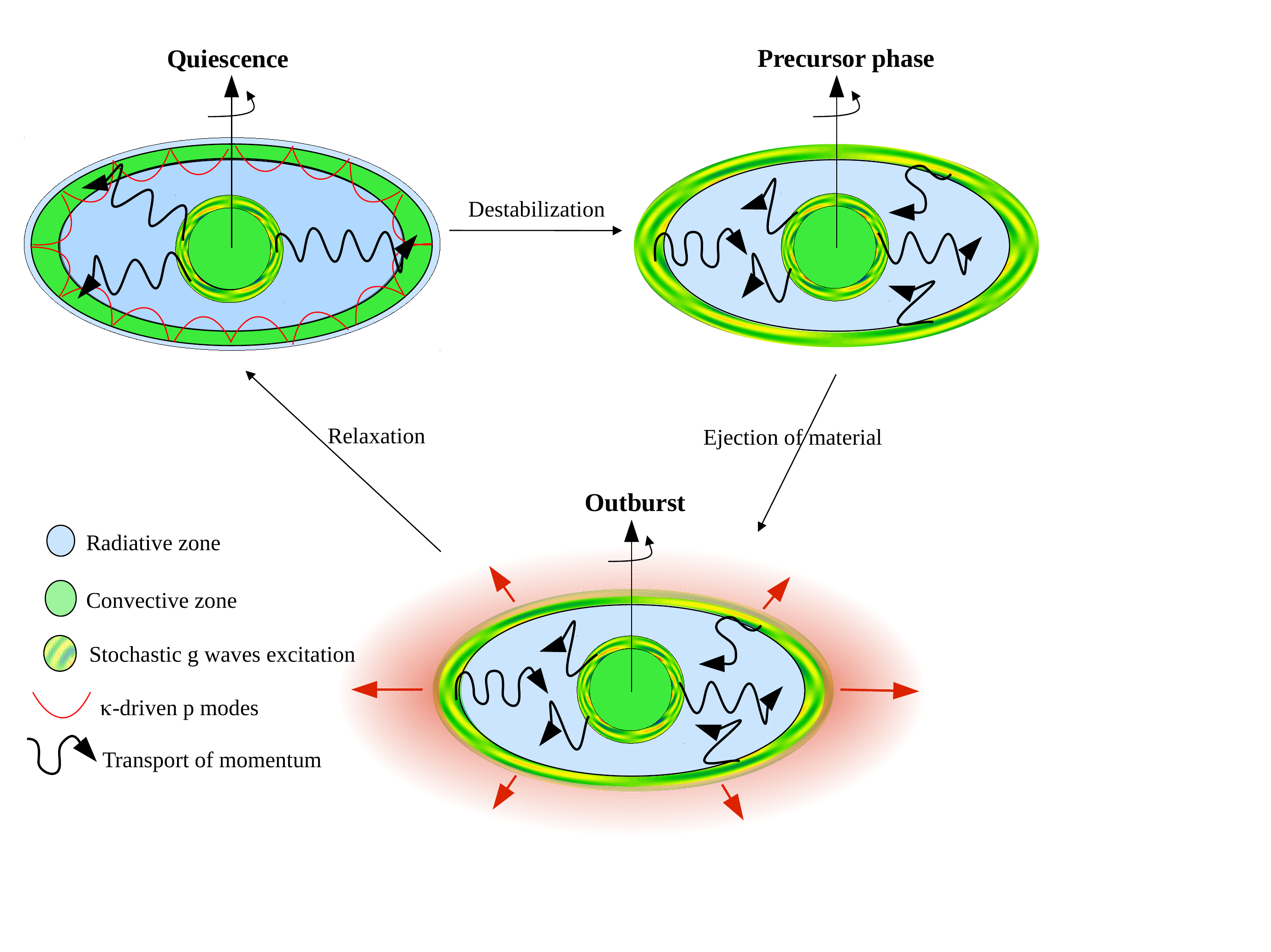}}
\caption[]{Sketch of the Be outburst of HD\,49330. In the quiescent phase, p modes are present and g modes stochastically excited at the interface between the convective core and the radiative envelope transport momentum towards the surface. Once enough momentum has accumulated, transient g modes get excited in the surface layer while p modes are destabilised and thus disappear. An ejection of material finally occurs. The star then relaxes back to a quiescent phase and
p modes reappear.}
\label{sketch}
\end{center}
\end{figure*}

First, the
modelling of the observed g modes, assuming they are excited by the $\kappa$
mechanism, requires a set of parameters with a much lower mass than the spectroscopically 
determined one. During the outburst, the star ejects material and its mass thus decreases. 
However, the mass ejected from the star during the
outburst is much lower than the difference between the mass determined from
observations and required in the model. Consequently, the mass loss during the
outburst cannot explain the excitation of g modes.

Second, a change in the radius of the star would influence the exact frequency
of the pulsations. \cite{huat}, however, did not observe changes in the
frequencies along the CoRoT light curve. They only witnessed changes in
amplitudes of the frequency peaks. Therefore, the change in radius during the
outburst cannot be the cause of the excitation of g modes.

Finally, the outburst lasted 73 days (including the precursor phase), which is
much shorter than the typical growth rate of $\kappa$-driven g modes of 10$^4$
times the oscillation period (see e.g. the middle panel of
Fig.~\ref{model_2D}). This indicates either that the observed g modes
were excited already before the precursor phase but not visible at the surface
of the star, or that these modes should not be considered as global oscillations
of the star.

\section{Alternative scenario: $\kappa$-driven p modes and stochastically-excited g waves}
\label{stochastic}

\subsection{Proposed scenario}

The results of our modelling 
show that p and g modes cannot be excited by the $\kappa$ mechanism at the same
time in HD\,49330 using the spectroscopically-determined stellar parameters. Moreover, 
the power spectrum of \cite{huat} shows broad frequency
groups rather than sharp frequency peaks around 1-2 c~d$^{-1}$ (see e.g. the bottom
panel of Fig.~\ref{model_2D}). Considering the above arguments, we propose that 
what \cite{huat} and \cite{floquet} attributed to $\kappa$-driven g modes are rather 
stochastic g waves. 

\begin{figure*}[!ht]
\begin{center}
\resizebox{!}{4.6cm}{\includegraphics[clip]{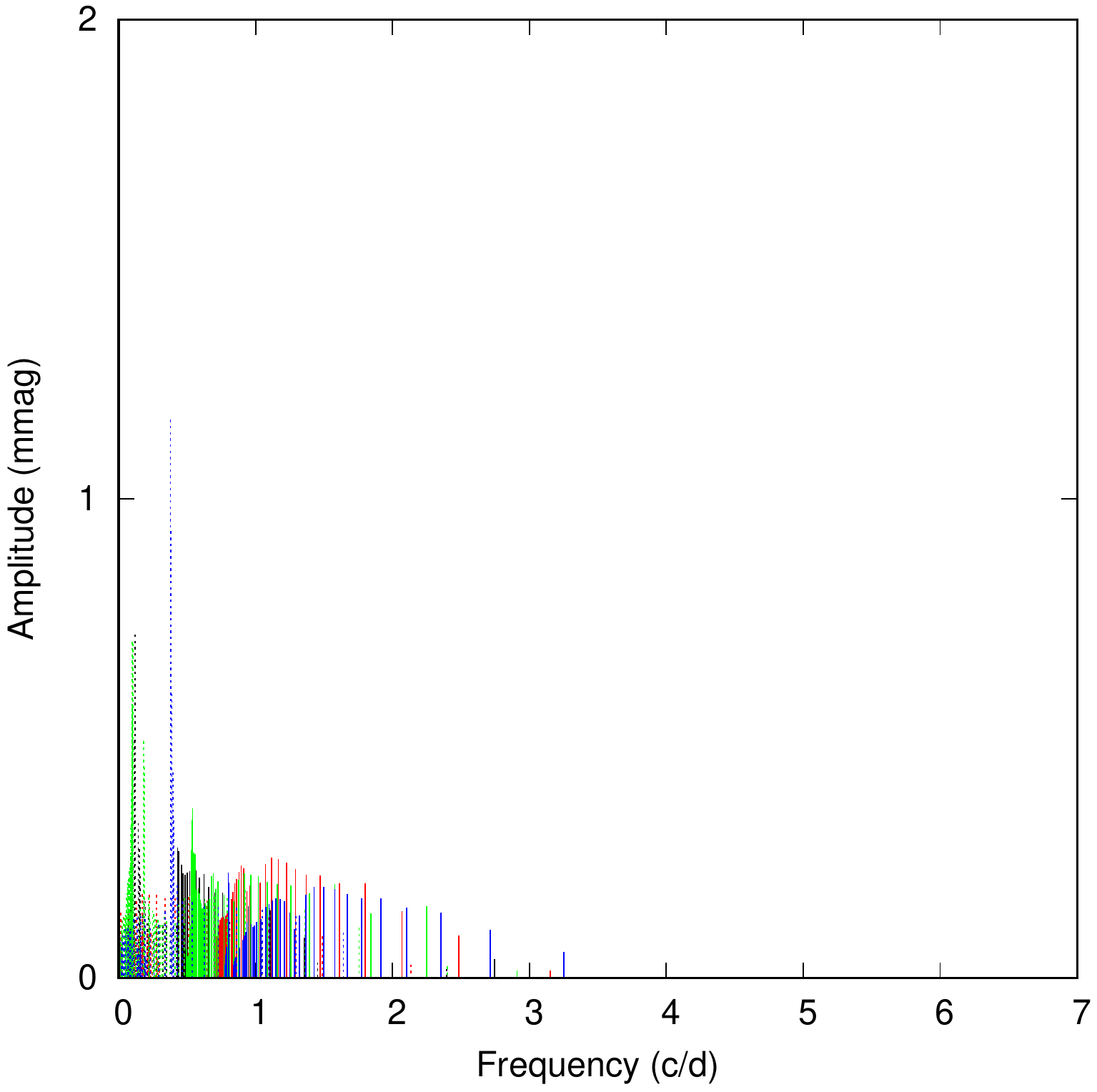}}
\resizebox{!}{4.6cm}{\includegraphics[clip]{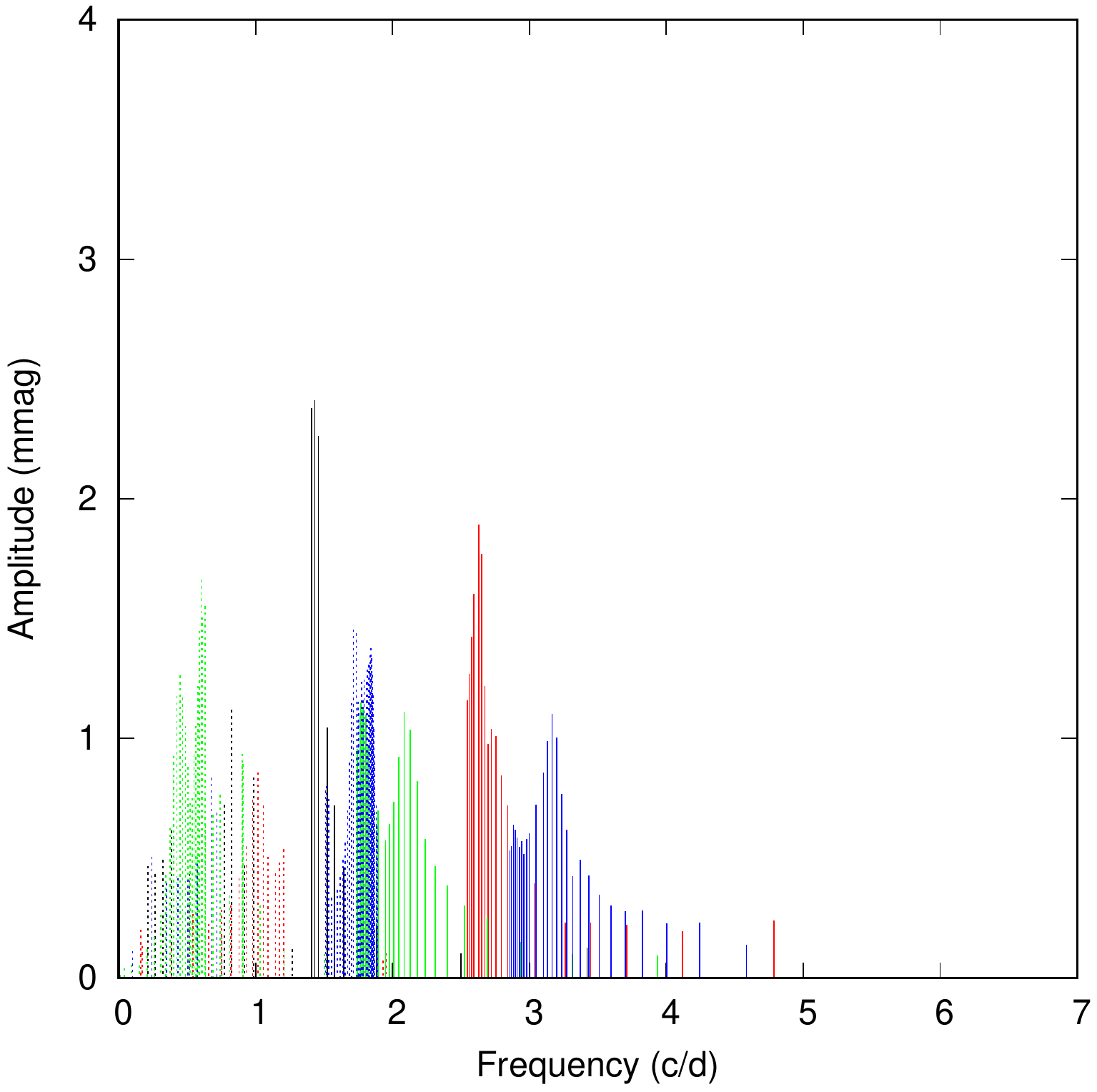}}
\resizebox{!}{4.6cm}{\includegraphics[clip]{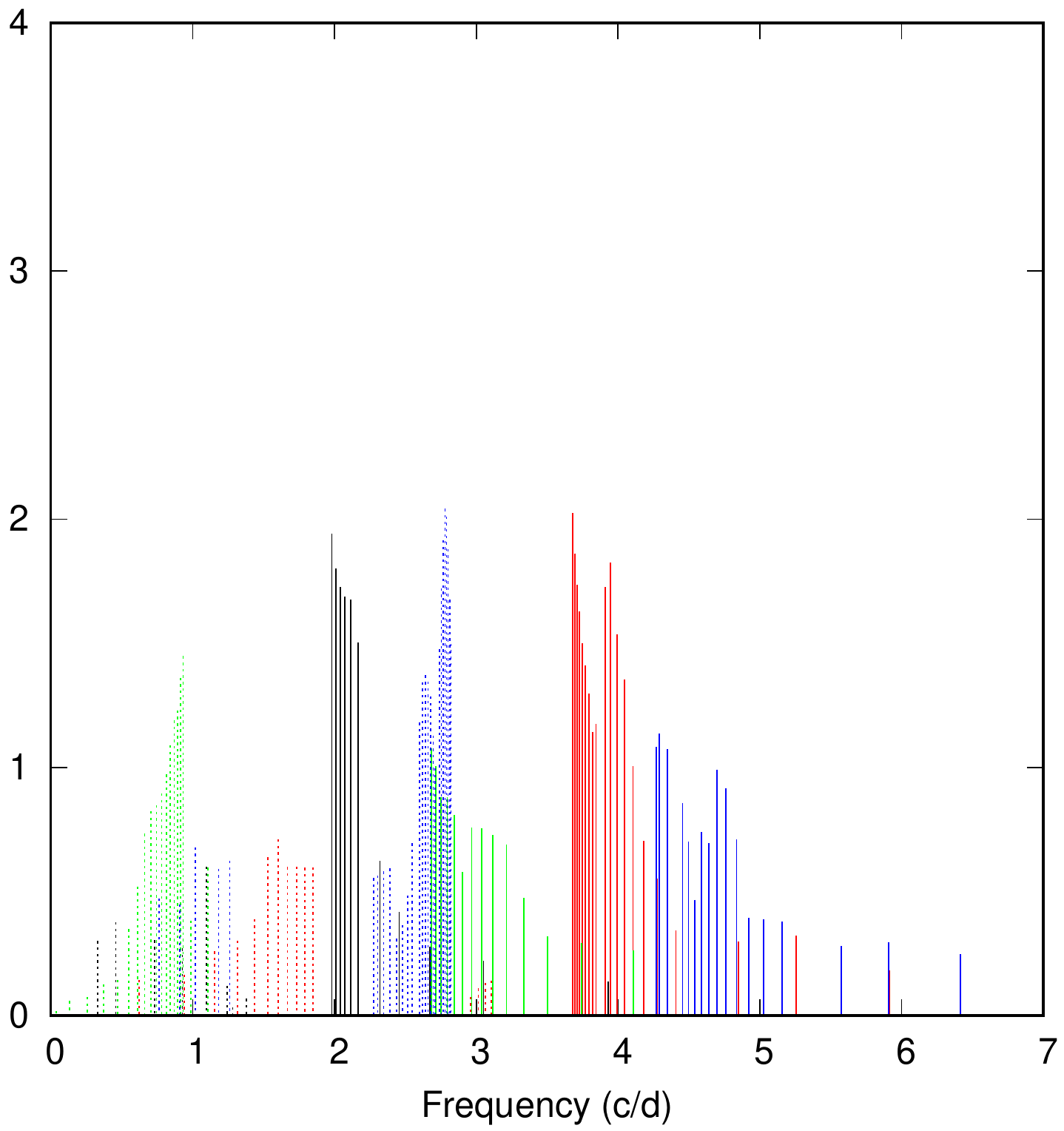}}
\resizebox{!}{4.6cm}{\includegraphics[clip]{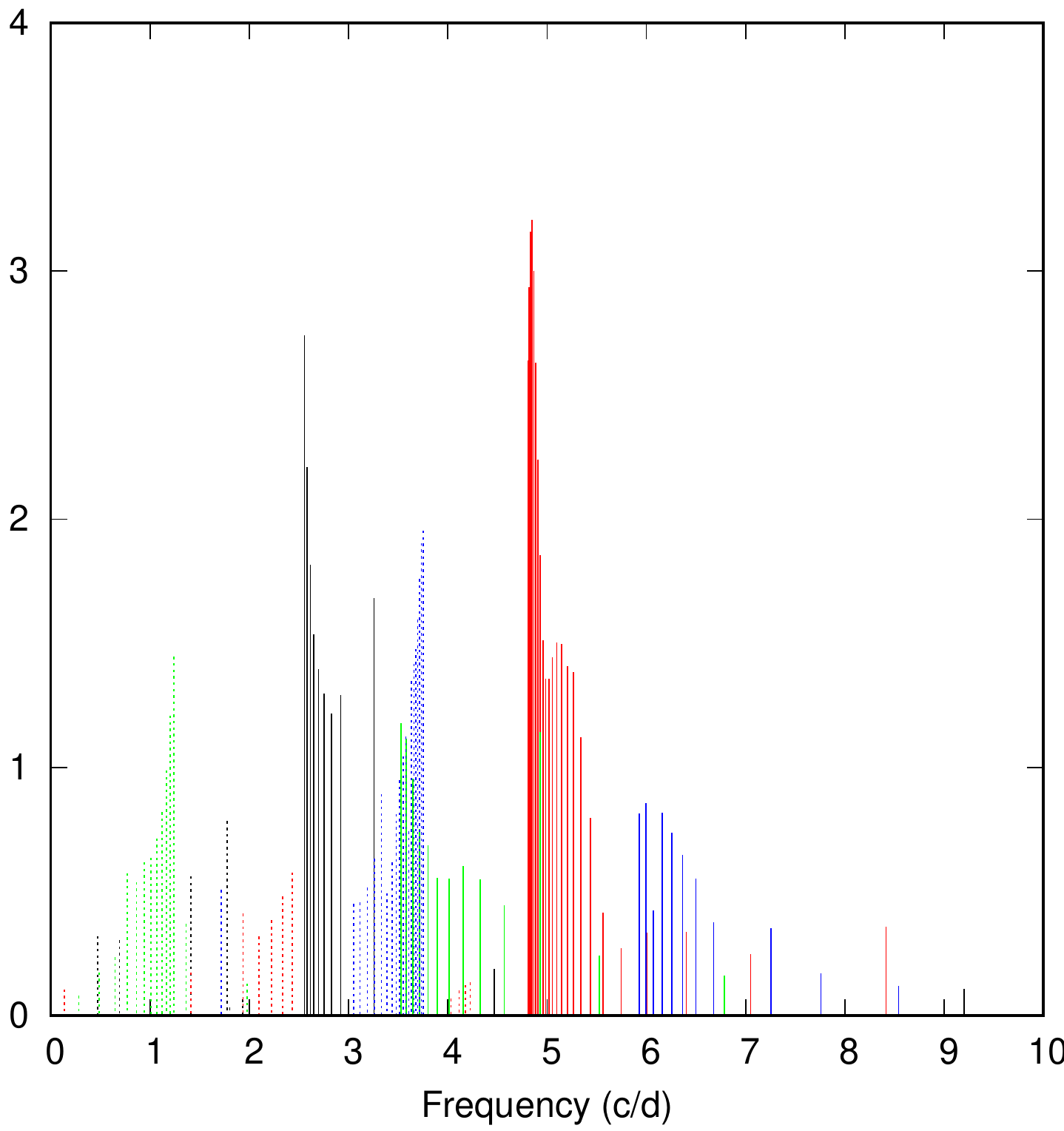}}
\caption[]{Amplitude of the stochastically excited g and r modes for various rotation rates: $\overline\Omega$=0.1 (left), $\overline\Omega$=0.4 (middle left), $\overline\Omega$=0.6 (middle right), and $\overline\Omega$=0.8 (right). The black lines represent l=|m|=1, red and green lines represent l=2 with |m|=2 and |m|=1 respectively, and blue lines represent l=3, |m|=2 modes. The solid and dotted lines are respectively for prograde and retrograde modes. The left panel has a different y-axis range to better see the low amplitudes.}
\label{figstocha}
\end{center}
\end{figure*}

During the quiet phase, stochastic gravity waves can be excited in the
convective core \citep{Browningetal2004,rogers2013,augustson2016}. These waves transport angular 
momentum from the core to the surface \citep{lee2014,rogers2015}. When enough angular momentum has 
accumulated in the outer layers of the star, these layers become unstable and emit 
transient gravity waves. The surface
layers reach the critical velocity, in particular at the equator where the
rotation was already the closest to critical. The destabilisation of the surface
layers thus ignites the outburst and breaks the cavity where the p modes were
propagating. This explains both the disappearance of the p modes during the
precursor and outburst phase of HD\,49330 and the ejection of material from the
surface into the disk, meaning the occurrence of the outburst, as observed by
\cite{huat} with CoRoT. Relaxation then occurs, recreating the cavity and
letting the p modes reappear while the surface transient g waves disappear. 

This outburst phenomenon will reoccur whenever the g modes stochastically excited in 
the core have transported a sufficient amount of angular momentum to the outer layers 
of the star. Both the stochastically excited g 
modes from the core and the transient g waves from the outer layers could be observed 
if their amplitude is high enough. A sketch of this
scenario is presented in Fig.~\ref{sketch}.

Below we test this scenario by modelling stochastically excited gravito-inertial
modes with the non-adiabatic Tohoku code modified for this purpose.

\subsection{Stochastic excitation of g waves}

\subsubsection{Theoretical background}

Convective regions, such as convective cores and subsurface convection zones in
massive stars, are able to stochastically excite oscillation modes
\citep{Cantielloetal2009,Belkacemetal2010} and particularly gravity modes
\citep{Samadietal2010}. The latter become gravito-inertial modes in fast
rotators such as HD\,49330 because of the action of the Coriolis acceleration
\citep[see e.g.][]{LeeSaio1997,DR2000,Mathis2009,Ballotetal2010}. This
excitation is now observed both in realistic numerical simulations of convective
cores surrounded by a stably stratified radiative envelope
\citep[see][]{Browningetal2004,rogers2013,augustson2016,Edelmannetal2019} and in asteroseismic data \citep{neinerStocha,aerts2015}.
Gravito-inertial waves are excited through their couplings with the volumetric
turbulence in the bulk of convective regions (where purely gravity modes in a
slowly rotating star are evanescent while gravito-inertial modes in a rapidly
rotating star become inertial) and by the impact of structured turbulent plumes
at the interfaces between convective and radiative regions. As a first step, we
choose to focus on the volumetric stochastic excitation. 

In this context, \cite{Belkacemetal2009} derived the general formalism to
treat the stochastic excitation of non-radial modes in rotating stars taking
into account the Coriolis acceleration. Their method is based on the results
first obtained by \cite{Samadietal2001} for radial p modes, which have been
extended to non-radial modes in a non-rotating star by \cite{Belkacemetal2008}. However, this formalism has been applied only to slowly rotating stars. Therefore we revisit it in the case of rapidly rotating stars, still ignoring centrifugal acceleration since \cite{Ballotetal2010} demonstrated that it can be neglected for g modes given the relatively weak distortion of the excitation region. The details of our new equations are provided in Appendix~\ref{appendix}. The main differences with \cite{Belkacemetal2009} are related to the strong action of the Coriolis acceleration. First, the expansion of the wave displacement now involves several spherical harmonics (see Eq. \ref{displacement}). Next, the Coriolis acceleration modifies the orthogonality relationships for eigenmodes \citep{Schenketal2002}. The normalisation of the amplitude becomes therefore more complex (see Eq. \ref{normalisation}). In this work, we still consider a simplified model for the convective turbulent source term, where the turbulence is assumed to be statistically stationary, incompressible, homogeneous, and isotropic, as in \cite{Samadietal2010}. These assumptions are assumed as a first step because of the lack of available robust prescriptions for convective rotating turbulence. A better model of turbulence would require specific developments which are beyond the scope of this work (we refer the reader to the Appendix~\ref{turb} and \cite{MNT2014} for a detailed discussion).

\subsubsection{Tohoku models with stochastically excited g modes}

We calculated seismic models of a $M$=15 M$\odot$ mass star with X = 0.7 and Z = 0.02 using the formalism described above and in Appendix~\ref{appendix} for $\overline{\Omega}$ = 0.1, 0.3, 0.4, 0.5, 0.6, and 0.8. We use series expansion for the perturbations of rotating stars. For the stochastic excitation of the low frequency modes, we compute $\nabla - \nabla_{ad}$ for the core using the usual mixing length theory with physical quantities of a ZAMS model. It is the first time that such models are computed for a rapidly rotating hot star. 

\begin{figure}[!ht]
\begin{center}
\resizebox{\hsize}{!}{\includegraphics[clip]{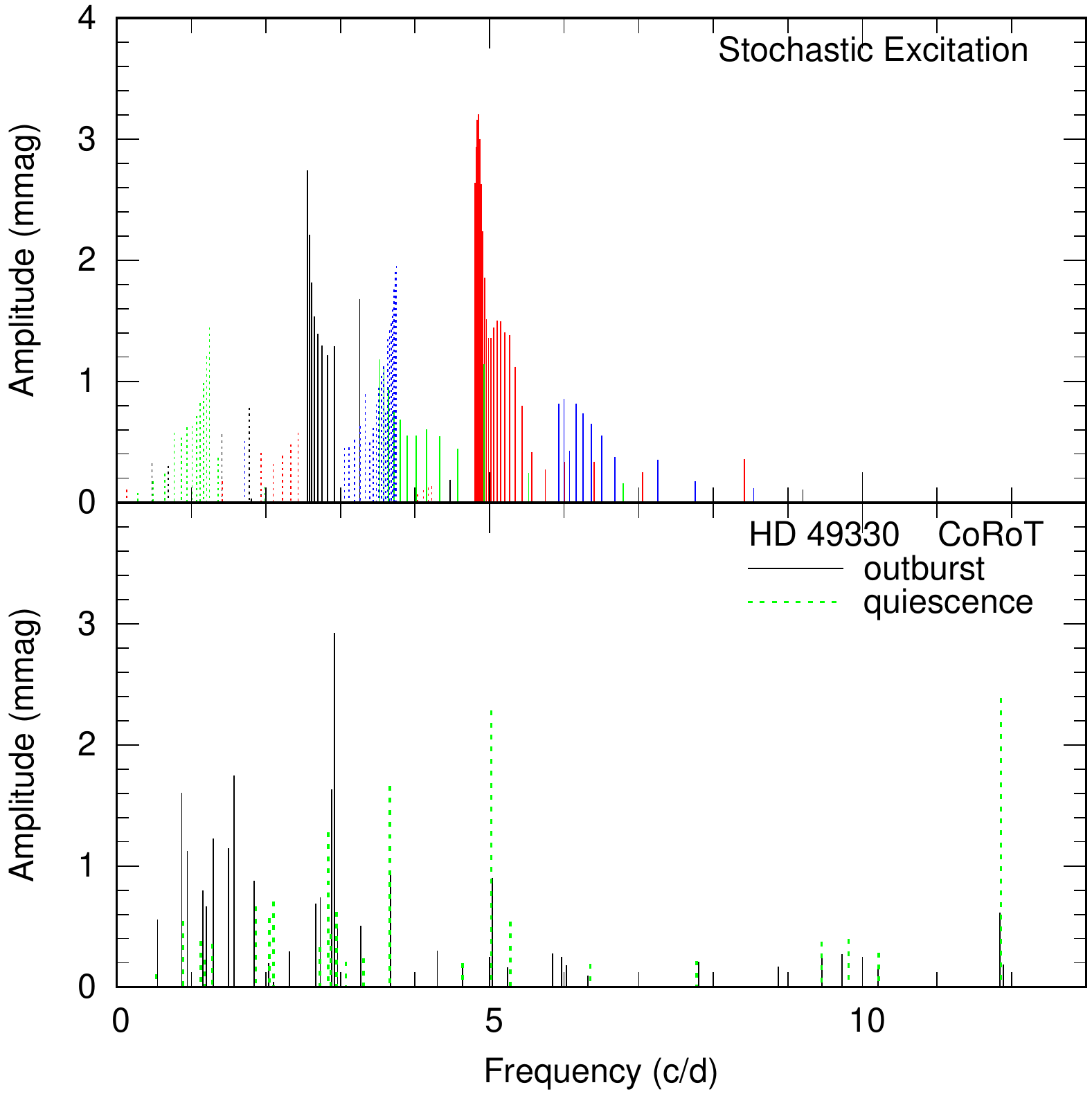}}
\caption[]{Top: Same as Fig.~\ref{figstocha} for $\overline\Omega$=0.8. Bottom: Statistically significant observed frequencies of HD\,49330 during and outside of the outburst phase as detected by \cite{huat}.}
\label{stochaobs}
\end{center}
\end{figure}

We find that the stochastic excitation of g and r modes occurs in this 15 M$\odot$ star model and thus the mass of HD\,49330 (M=14.4 M$\odot$) is not a concern anymore to produce the observed frequencies with this excitation mechanism (contrary to $\kappa$-driven modes).

Fig.\ref{figstocha} shows the variation in stellar surface magnitude produced by these waves as a function of their frequency for $\overline{\Omega}$ = 0.1, 0.4, 0.6, and 0.8. Since the star is considered spherical here, $\overline{\Omega}$ = 0.1, 0.4, 0.6, and 0.8 would correspond to 0.12, 0.46, 0.69 and 0.92 at the equator if the star was deformed as in the 2D {\sc ROTORC} model shown above. The amplitude is computed from the maximum value of the luminosity variation at the surface induced by the stochastically excited waves. We refer the reader to Appendix A of \cite{leesaio1993} for the equations solved for non-adiabatic oscillations that allow the computation of the luminosity variation in the code.

For the figure, since for rapidly rotating stars mode identification becomes unclear particularly for retrograde modes, we consider non-adiabatic pulsations for which the kinetic energy $l= |m|$ even parity and $l=|m|+1$ odd parity modes is above 50\% of the total kinetic energy of the sum of all components constituting an eigenmode for a given m. We include g and r modes only for $|m|=1$ and 2 with $l=|m|$ and $|m|+1$. We find that the amplitude of the even parity, prograde modes is the strongest. 

The computed amplitude for $\overline{\Omega} \geq 0.4$ is similar to the amplitudes detected with CoRoT, which confirms that we likely see stochastically excited gravito-inertial modes in HD\,49330. In fact, the modelled amplitude is slightly higher than the amplitude observed for HD\,49330 (see Fig.~\ref{stochaobs}, which shows in the bottom panel only the detected frequencies considered statistically significant, that is with a S/N above 4, while many more lower amplitude frequencies are present in the data). This may be because we considered a $M$=15 M$\odot$ model, while a $M$=14.4 M$\odot$ star such as HD\,49330 would have a lower luminosity. In addition, our model is not complete since only the action of Reynolds stresses associated to convective small scales has been considered while large scale convective structures such as turbulent plumes likely also contribute to the stochastic excitation of gravito-inertial waves \citep{schatzman1993, rogers2013, alvan2014, pincon2016}, and the assumed turbulence source term is very simple (see Appendix~\ref{turb}).

Nevertheless, the models show that an increasing rotation rate modifies the amplitude, allowing certain frequencies to be excited with higher amplitudes for rapid rotators. This is the result of a better coupling between gravito-inertial waves and the convective forcing, in particular in the sub-inertial regime in which gravito-inertial waves become propagative inertial waves in convective regions. Such a behaviour has been observed in numerical simulations \citep{rogers2013} and predicted theoretically \citep{MNT2014}.

However, the variation in the amplitude of some of the modes in our models is not a monotonic function of the rotation rate. For example, the amplitude peaks for $\overline{\Omega}$=0.4 and 0.8 for l=|m|=1 modes, while it increases with rotation rates for l=|m|=2 modes. This is because both the damping rate of modes and the injection of energy are not monotonic. Moreover, above $\overline{\Omega}$=0.4, the fact that we ignored the deformation from a spherical star owing to centrifugal acceleration in the computation of waves \citep{Ballotetal2010,MP2019} as well as the increase in the size of the core observed in rapidly rotating Be stars \citep{neinerMixing} may have more impact.

Finally, as expected, we find that more frequencies are excited when the star rotates rapidly \citep{rieutord2009}, and most g modes are in the sub-inertial regime. This forest of stochastically excited g modes with higher amplitudes on average could help to transport angular momentum in rapidly rotating stars.

\subsection{Transport of angular momentum}\label{transport}

The aim of this subsection is to test the correlation between the transport of
angular momentum by gravito-inertial waves, which are stochastically excited
inside the star, and Be outbursts. 

\begin{figure*}[!ht]
\begin{center}
\resizebox{!}{4.2cm}{\includegraphics[clip]{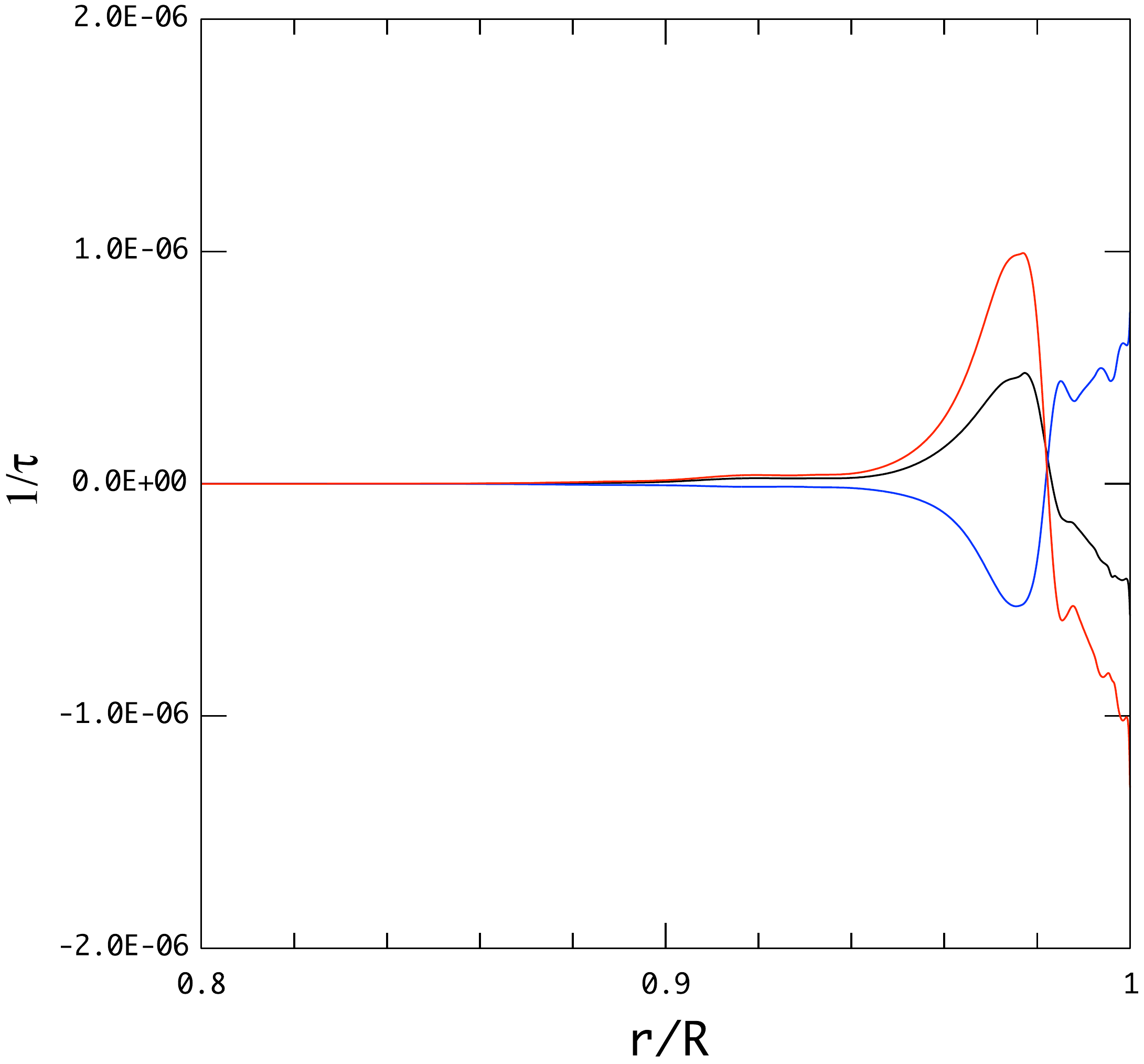}}
\resizebox{!}{4.2cm}{\includegraphics[clip]{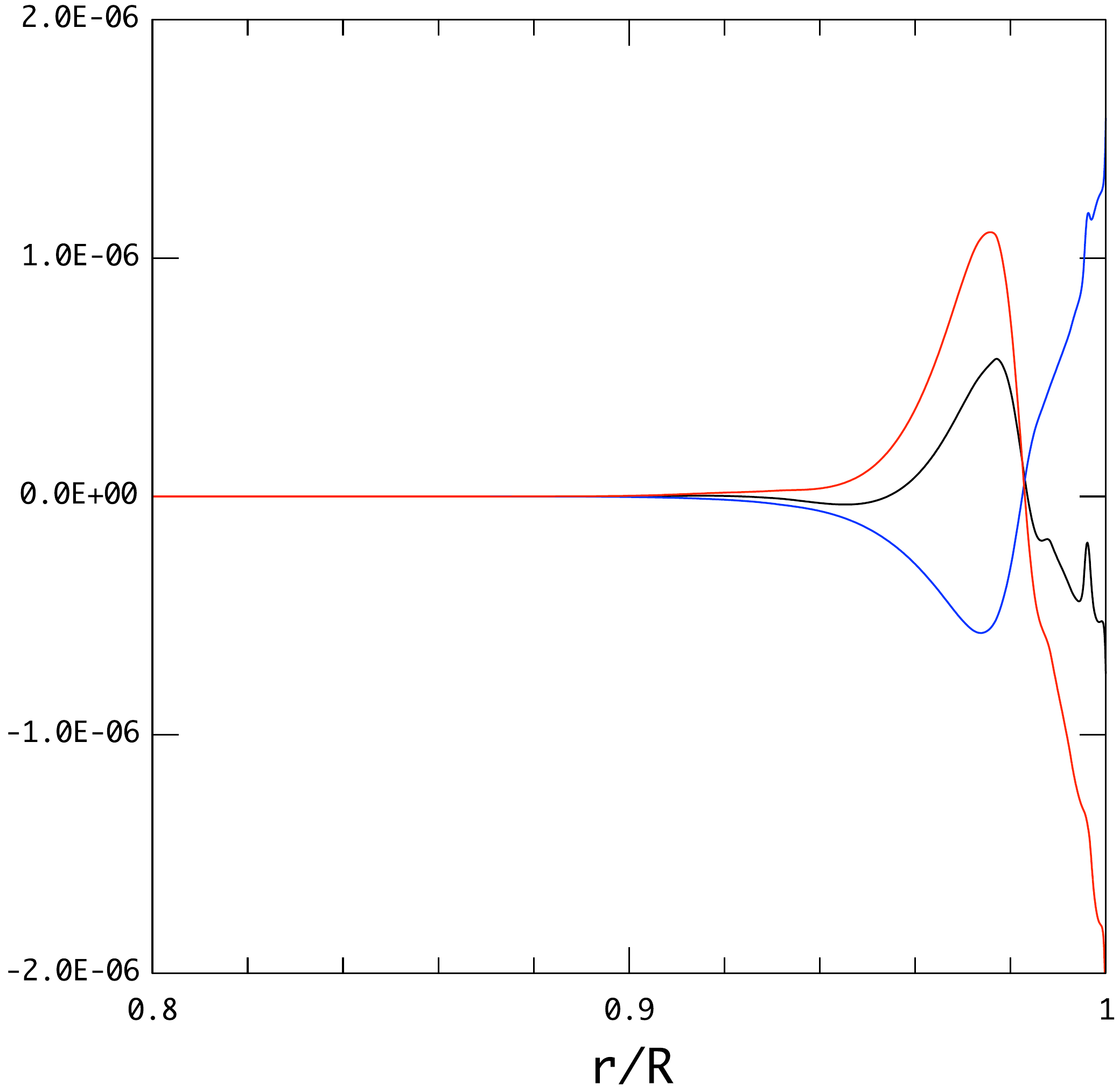}}
\resizebox{!}{4.2cm}{\includegraphics[clip]{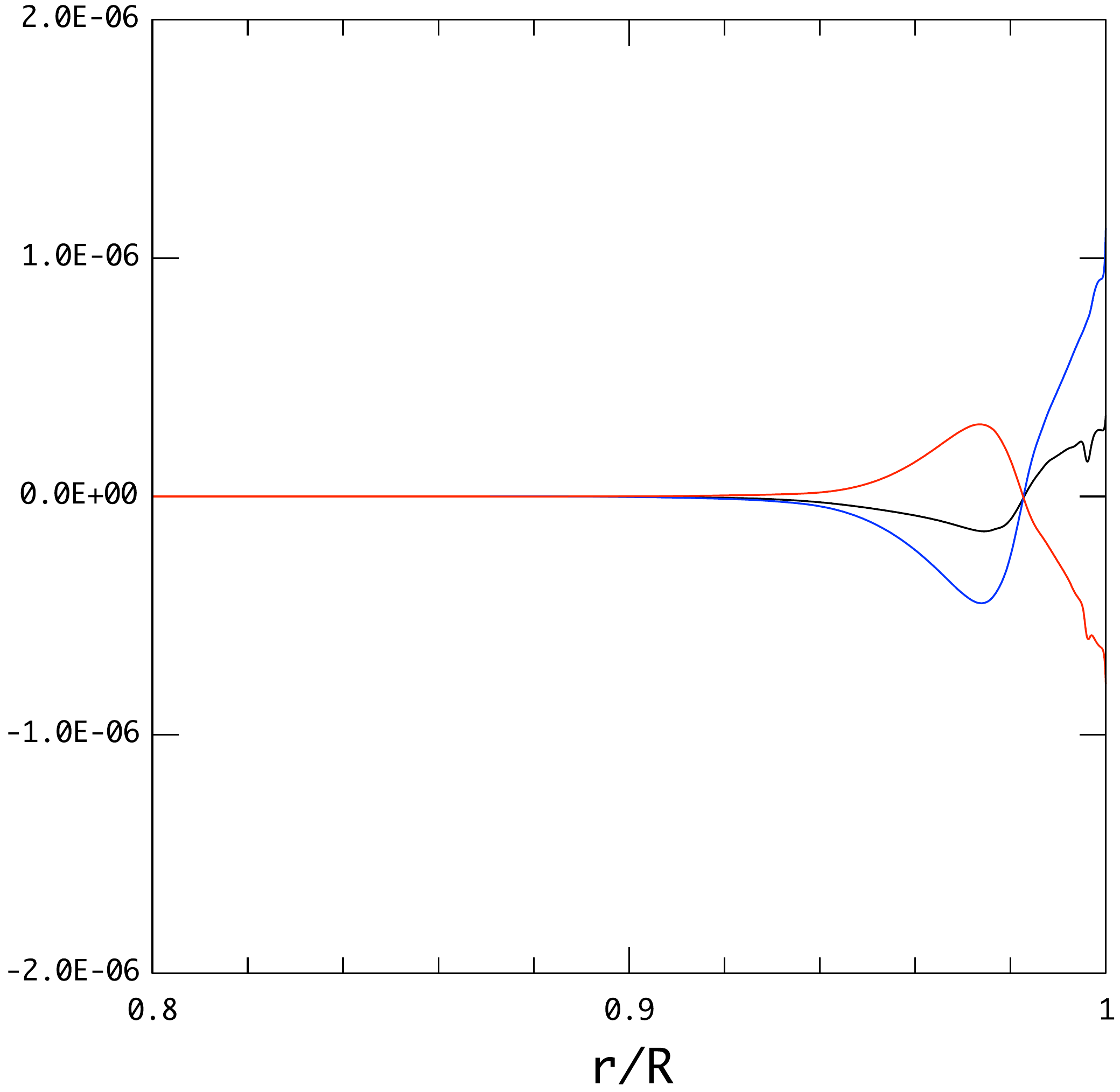}}
\resizebox{!}{4.2cm}{\includegraphics[clip]{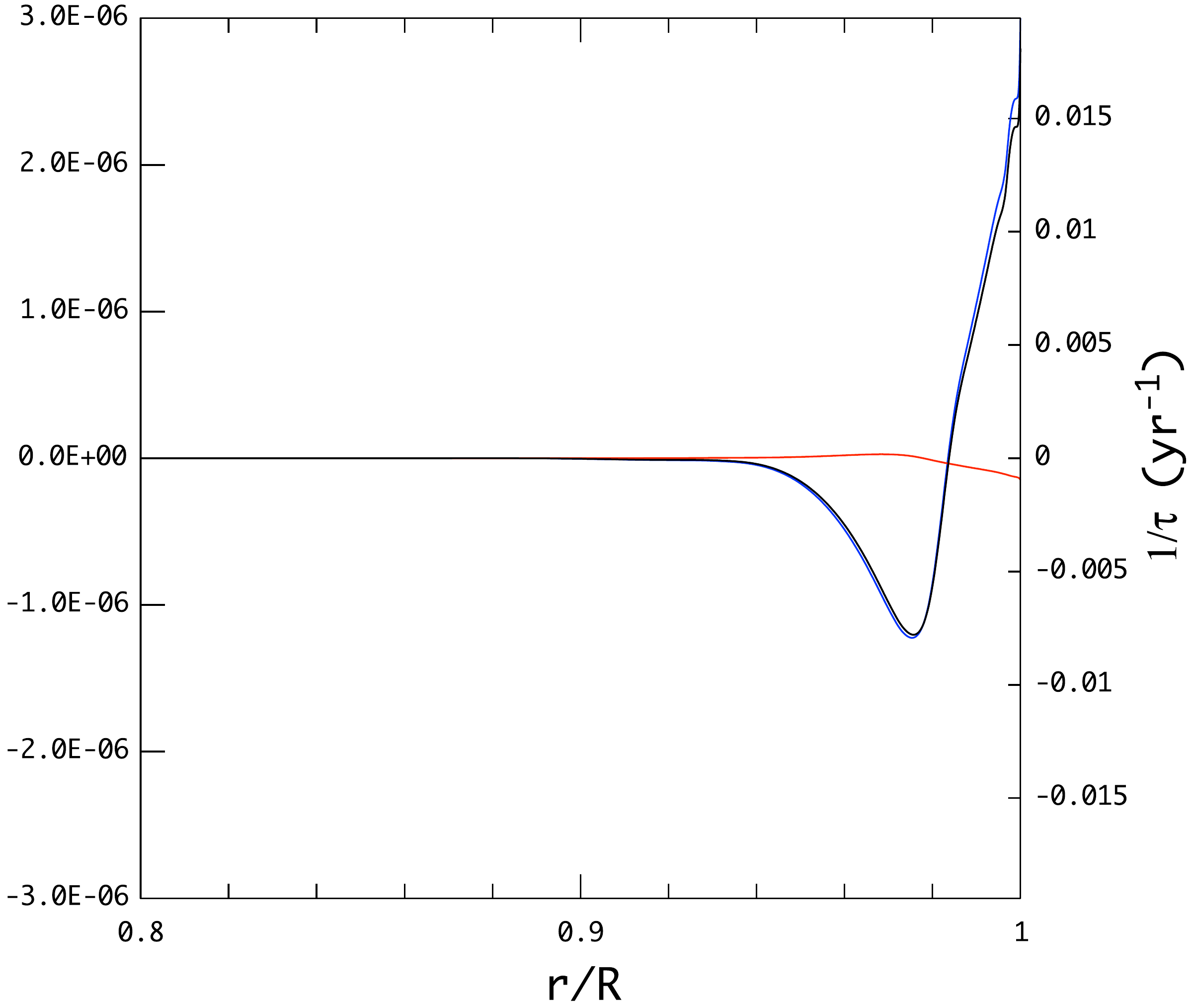}}
\caption[]{
Inverse of the timescale of retrograde (red line), prograde (blue line) and the sum of the two (black line), g and r modes for a $M$=15 $M_{\odot}$ star with $\overline\Omega$ = 0.1 (left), 0.4 (middle left), 0.6 (middle right) and 0.8 (right), where $\tau$ is normalised to $\sqrt{GM/R^3}$. The value of 1/$\tau$ is proportional to the angular momentum transport. Note that the right panel (fastest rotator) has a different y-axis range to accommodate the larger 1/$\tau$ values and shows an extra y-axis on the right side in years$^{-1}$.}
\label{figtransport}
\end{center}
\end{figure*}

First, we recall that the idea that non-radial oscillations excited in massive stars are able to efficiently transport angular momentum through their damping by thermal diffusion below the stellar surface and that they may impel the surface of a Be star to reach its critical velocity has already been proposed by
\cite{ando1986}, \cite{lee_saio}, and \cite{lee2006}. However, in these previous
works, gravito-inertial waves (and r modes) were excited by the
$\kappa$-mechanism. In the case of HD\,49330, we propose that gravito-inertial
waves are stochastically excited in turbulent convective regions.
\cite{pantillon2007} and \cite{lee2014} demonstrated that such waves are also able to transport
angular momentum. 

Following the method of \cite{lee2012} \citep[see
also][]{grimshaw1984,lee2006,mathis_debrye}, we consequently study the transport
of angular momentum induced by stochastically excited gravito-inertial waves. We recall that here we ignore the centrifugal acceleration (i.e. the star is considered spherical) both for the stellar structure model and the pulsations computations (contrary to the $\kappa$-driven modes computed in Sect.~\ref{sect_kappa}).
Following this method, we calculate the acceleration of the mean azimuthal flow
on an isobar. We consider the horizontally averaged equation describing the
interaction between the wave and the mean flow, given by:

\begin{equation}
{\overline\rho}\frac{{\rm d}\ell}{{\rm d}t}=-\sum_{\alpha}\left\{\frac{1}{r^2}\partial_{r}\left(r^2 \overline{{F}_{\alpha}^{\rm AM;V}}\right)+\overline{\left<\rho_{\alpha}^{'}\partial_{\varphi}\Phi_{\alpha}^{'}\right>}\right\},
\end{equation}

\noindent where the subscript $\alpha$ identifies the excited mode, 
$\left<\cdot\!\cdot\!\cdot\right>=\frac{1}{2\pi}\int_{0}^{2\pi}\cdot\!\cdot\!\cdot\,{\rm
d}\varphi$ and
$\overline{\cdot\!\cdot\!\cdot}=\frac{1}{\pi}\int_{0}^{\pi}\cdot\!\cdot\!\cdot\,\sin\theta{\rm
d}\theta$. Here, $\ell=\overline{\left<r^2\sin^2\theta\,\Omega+r\sin\theta\,{\overline v}_{\varphi}\right>}$  is
the specific angular momentum around the rotation axis, where $\Omega$ is the
considered uniform rotation, and ${\overline{v_{\varphi}}}/{r\sin\theta}$ is the
induced differential rotation. Each scalar field ($X$) has been expanded as
$X={\overline X}+X^{'}$, where $\overline X$ is the mean value on an isobar and
$X^{'}$ is the modal fluctuation. The vertical flux of angular momentum is given
by

\begin{equation}
{F}_{\alpha}^{\rm AM;V}={\overline\rho}r\sin\theta\left<v_{r}^{'}\left(v_{\varphi}^{'}+2\Omega\cos\theta\xi_{\theta}\right)\right>,
\end{equation}

\noindent where we recall that, for modes calculated taking only the given
uniform rotation into account, ${\vec v}=i\left(\omega_{0}+m\Omega\right)\vec\xi$.

We define the associated timescale for the transport of angular momentum $\tau$ as:

\begin{equation}
\frac{1}{\tau}=\frac{{\rm d}\ln \ell}{{\rm d}t}.
\end{equation}

Figure~\ref{figtransport} shows the inverse of the timescale of stochastically excited prograde and retrograde g and r modes for slowly and rapidly rotating stars with $M$=15 $M_\odot$. For a sufficiently rapid rotation rate, angular momentum is transported just below the surface. As expected, retrograde modes extract angular momentum, whereas prograde modes deposit angular momentum at the surface, leading respectively to a deceleration and acceleration of the surface rotation. Below the surface (at about r = 0.975 R), the modes stochastically excited in the core experience an additional $\kappa$ excitation at the position of the opacity bump, which creates the reversal of sign in the angular momentum flux (see Fig.~\ref{figtransport}). In the case of a slowly rotating star, the net transport (sum of deposition and extraction) is very low. In the case of a rapidly rotating star, however, most of the retrograde modes are $r$ modes, for which the pressure and density perturbations are small so that their angular momentum transfer is inefficient. Therefore, the prograde modes deposit much more momentum than is extracted by the retrograde modes at the surface. Thus, the net angular momentum transport by stochastically excited modes is large and clearly increases with the rotation rate at the surface. These modelling results confirm the prediction by \cite{MNT2014} and \cite{lee2014}.

In addition, the minimum timescale $\tau$ is attained around r = 0.975 R and is of the order of $8.3\times10^5$  ($1/\tau=1.2\times10^{-6}$) for the sum of prograde and retrograde modes in a rapidly rotating star. Since $\Omega_c$ is about $1.5\times10^{-4} s^{-1}$ in this model, the acceleration timescale is of the order of 125 years. This timescale would correspond to the time needed for the first outburst to occur, that is for the required angular momentum to be transported from the core to the surface and accumulated there.

A characteristic recurrence timescale of outbursts defined as
\begin{equation}
\frac{1}{\tau'}=\frac{{\rm d}\ln\overline{\left<r^2\sin^2\theta\,\Omega_c-\left(r^2\sin^2\theta\,\Omega+r\sin\theta\,{\overline v}_{\varphi}\right)\right>}}{{\rm d}t}
\end{equation}
can also be computed. In the case of a shellular rotation that depends only on the radial coordinate, it becomes
\begin{equation}
\tau'=\tau\vert{\overline\Omega}^{\,-1}-1\vert.    
\end{equation}
It evaluates how much time is needed to reach again critical velocity at the surface after a first outburst, assuming that the transport triggered by waves occurs continuously; if ${\overline\Omega}\rightarrow 1$ then $\tau'\rightarrow 0$, meaning if the star rotates at critical velocity it looses mass all the time, while if ${\overline\Omega}\rightarrow 0$ then  $\tau'\rightarrow\infty$, meaning if a star does not rotate no outburst occurs. $\overline\Omega$=0.8 means $\overline\Omega$=0.92 in the non-spherical case. We then predict that an outburst should occur every $\sim$11 years in HD\,49330. Photometric observations by ASAS-3 have shown that an outburst occurred around December 2002, while the outburst observed by CoRoT occurred in February 2008 \citep{huat}. The time difference between these two outbursts is about half but still similar to the recurrence timescale computed here. However, a more regular monitoring of the lightcurve of HD\,49330 is needed to confirm the observed recurrence of its outbursts.

\subsection{Outburst}\label{outburst}

The accumulation of angular momentum near the surface destabilises the outer layers of the star. This situation is similar to local forces and heating or cooling, such as a turbulent layer, wind, or convective structure in the terrestrial atmosphere, which generate gravity waves \citep[e.g.][]{vadas2001,townsend1965}. Realistic models for such excitation have been developed in atmospheric physics but never applied in the context of hot stars. Nevertheless, by analogy, we expect the perturbation excited in the outer layers of HD\,49330 to produce local mechanical and thermal forcing that generate transient g waves. Moreover, once the amount of angular momentum deposited in the surface layers of the star impels it to the breakup velocity, the star starts loosing mass, meaning the outburst occurs.

Both the g modes from the core and the g waves from the outer layers could be observable. In the CoRoT observations of HD\,49330, g waves are mostly visible during the precursor and outburst phases, therefore they must be the transient g waves from the outer layers. Nevertheless, some peaks around 3 and 3.6 c~d$^{-1}$ seem to also be present during quieter phases \citep[see Fig.~4 in][]{huat} and could originate in the core. 

\section{Conclusions}
\label{conclusion}

\subsection{Summary of the results}

Here, we summarise the main findings of this work:
\begin{enumerate}
\item The treatment of rotational deformation is very important for high-frequency p modes, because the excitation of these modes seems to occur only when the rotation is nearly critical. Consequently, Chandrasekhar-Milne's expansion is not a good approximation to model p modes of rapidly rotating early B stars, such as early Be stars or rapidly rotating $\beta$ Cephei stars. Using a 2D structure model is essential in this case.
\item Contrary to what is usually assumed, Be stars seem to host not only $\kappa$-driven pulsation modes, but also stochastically excited g waves produced continuously in the convective core as well as in a transient way in the outer layers of the star during an outburst. A direct detection of stochastically excited gravito-inertial modes has been obtained by \cite{neinerStocha} in the Be star HD\,51452. This should be carefully taken into account for future interpretations of seismic data of Be stars. Indeed, while for early Be stars, it is clear that $\kappa$-driven g modes cannot be excited and thus that any observed g mode is stochastically excited \citep{neinerStocha}, for late Be stars g modes or waves can be excited both by the $\kappa$ mechanism and stochastically.
\item The amplitude of stochastically excited modes increases with rotation rate on average. However, we find that the amplitude of some modes does not vary monotonically with $\overline{\Omega}$. This is likely related to the competition between the injection of energy and damping rates, which are not monotonic functions of the rotation rate. The l=2, m=-2 modes, however, have a clear increase in magnitude with rotation rate and contribute significantly to the transport of angular momentum. Those modes are the ones most commonly observed in Be stars \citep{rivinius2003}.
\item The transport of angular momentum by stochastically excited modes clearly increases with rotation rate and becomes very strong for stars rotating close to critical. This is because prograde modes dominate retrograde modes. In particular, the l=2, m=-2 modes have a strong amplitude when the star rotates fast.
\item The recurrence timescale of outbursts produced by this angular momentum transport mechanism is of the order of 11 years for HD\,49330, which is similar to the outbursts occurrence observed in this star.
\item The outburst in HD\,49330 is due, in addition to the almost critical velocity of the star, to the transport of angular momentum to the surface layers by stochastic gravity waves generated in the convective core. The accumulation of angular momentum in the surface layers is at the origin of the destabilisation of the surface layers, of the production of transient g waves in these layers, and of the outburst. This explains the correlation between the variation in the amplitude of the pulsation frequencies observed in the CoRoT data and the occurrence of the outburst. A similar scenario could occur in other pulsating Be stars, thus providing an explanation for the long-standing question of the origin of Be outbursts and disks, at least for some Be stars.
\end{enumerate}

\subsection{Limitations and future work}

This work is a first attempt at coherently modelling stochastically excited gravito-inertial pulsation modes, and the transport of angular momentum they induce, in a rapidly rotating Be star. The formalism we developed here for Be stars can be applied to other rapidly rotating g-mode pulsators, such as $\gamma$\,Dor or SPB stars. 

However, this first attempt can be improved in the future. In particular, the model presented here for stochastically excited gravito-inertial modes does not include centrifugal acceleration, neither for the stellar structure model nor for the pulsation computation, contrary to the models of $\kappa$-driven modes that we present. Rotation is nevertheless partly taken into account via the Coriolis acceleration for both the stellar structure and oscillations. In the future, it would be interesting to use a 2D stellar structure to see how this may change the amplitude of the modes. Moreover, \cite{neinerMixing} showed that mixing induced by the penetrative movements at the bottom of the radiative envelope and by the secular hydrodynamical transport processes induced by the rotation in the envelope extend the size of the core of Be stars, which may increase stochastic excitation.

In addition, we used a standard description of turbulent convection in the core, which does not account for rapid rotation. Indeed, for a rapidly rotating star, convection becomes strongly anisotropic and the transfer of energy may be modified by rapid rotation. However, no robust prescription for these effects are available as of this writing. 
Furthermore, the formalism derived here for stochastic excitation focuses on the action of Reynolds stresses associated to convective small scales while large scale convective structures such as turbulent plumes likely also contribute to the stochastic excitation of gravito-inertial waves. 

Finally, we considered $\kappa$-driven p and g modes as well as stochastically excited gravito-inertial (g and r) modes. \cite{saio2018MNRAS} and \cite{saio2018arXiv} showed that r modes could also be excited mechanically in most rotating stars, including Be stars. In this case, even r modes of order $m$ appear as a group of frequencies just below $m$ times the rotation frequency in the Fourier spectrum of those stars. Such mechanically excited r modes can be a transient phenomenon at the surface and could therefore be an alternative explanation to the low frequencies observed during the outburst of HD\,49330, which will be studied in a separate paper (Saio et al. in prep.). 

\begin{acknowledgements}
The authors thank the referee for their comments, which have allowed us to improve the article. The CoRoT space mission, launched on December 27th 2006, has been developed and was operated by CNES, with the contribution of Austria, Belgium, Brazil, ESA (RSSD and Science Program), Germany and Spain. CN acknowledges fundings from the SIROCO ANR project, CNES, and PNPS. SM and KCA acknowledge support from the ERC SPIRE 647383 grant and PLATO CNES grant at CEA/DAp-AIM. This research has made use of the SIMBAD database operated at CDS, Strasbourg (France), and of NASA's Astrophysics Data System (ADS).
\end{acknowledgements}

\bibliographystyle{aa}
\bibliography{articles}

\begin{thebibliography}{59}
\expandafter\ifx\csname natexlab\endcsname\relax\def\natexlab#1{#1}\fi

\bibitem[{{Aerts} \& {Rogers}(2015)}]{aerts2015}
{Aerts}, C. \& {Rogers}, T.~M. 2015, \apjl, 806, L33

\bibitem[{{Alvan} {et~al.}(2014){Alvan}, {Brun}, \& {Mathis}}]{alvan2014}
{Alvan}, L., {Brun}, A.~S., \& {Mathis}, S. 2014, \aap, 565, A42

\bibitem[{{Ando}(1986)}]{ando1986}
{Ando}, H. 1986, \aap, 163, 97

\bibitem[{{Augustson} {et~al.}(2016){Augustson}, {Brun}, \&
  {Toomre}}]{augustson2016}
{Augustson}, K.~C., {Brun}, A.~S., \& {Toomre}, J. 2016, \apj, 829, 92

\bibitem[{{Augustson} \& {Mathis}(2019)}]{AugustsonMathis2019}
{Augustson}, K.~C. \& {Mathis}, S. 2019, \apj, 874, 83

\bibitem[{{Auvergne} {et~al.}(2009){Auvergne}, {Bodin}, {Boisnard}, {Buey},
  {Chaintreuil}, {Epstein}, {Jouret}, {Lam-Trong}, {Levacher}, {Magnan},
  {Perez}, {Plasson}, {Plesseria}, {Peter}, {Steller}, {Tiph{\`e}ne}, {Baglin},
  {Agogu{\'e}}, {Appourchaux}, {Barbet}, {Beaufort}, {Bellenger}, {Berlin},
  {Bernardi}, {Blouin}, {Boumier}, {Bonneau}, {Briet}, {Butler}, {Cautain},
  {Chiavassa}, {Costes}, {Cuvilho}, {Cunha-Parro}, {de Oliveira Fialho},
  {Decaudin}, {Defise}, {Djalal}, {Docclo}, {Drummond}, {Dupuis}, {Exil},
  {Faur{\'e}}, {Gaboriaud}, {Gamet}, {Gavalda}, {Grolleau}, {Gueguen},
  {Guivarc'h}, {Guterman}, {Hasiba}, {Huntzinger}, {Hustaix}, {Imbert},
  {Jeanville}, {Johlander}, {Jorda}, {Journoud}, {Karioty}, {Kerjean},
  {Lafond}, {Lapeyrere}, {Landiech}, {Larqu{\'e}}, {Laudet}, {Le Merrer},
  {Leporati}, {Leruyet}, {Levieuge}, {Llebaria}, {Martin}, {Mazy}, {Mesnager},
  {Michel}, {Moalic}, {Monjoin}, {Naudet}, {Neukirchner}, {Nguyen-Kim},
  {Ollivier}, {Orcesi}, {Ottacher}, {Oulali}, {Parisot}, {Perruchot},
  {Piacentino}, {Pinheiro da Silva}, {Platzer}, {Pontet}, {Pradines},
  {Quentin}, {Rohbeck}, {Rolland}, {Rollenhagen}, {Romagnan}, {Russ}, {Samadi},
  {Schmidt}, {Schwartz}, {Sebbag}, {Smit}, {Sunter}, {Tello}, {Toulouse},
  {Ulmer}, {Vandermarcq}, {Vergnault}, {Wallner}, {Waultier}, \&
  {Zanatta}}]{auvergne}
{Auvergne}, M., {Bodin}, P., {Boisnard}, L., {et~al.} 2009, \aap, 506, 411

\bibitem[{{Ballot} {et~al.}(2010){Ballot}, {Ligni{\`e}res}, {Reese}, \&
  {Rieutord}}]{Ballotetal2010}
{Ballot}, J., {Ligni{\`e}res}, F., {Reese}, D.~R., \& {Rieutord}, M. 2010,
  \aap, 518, A30

\bibitem[{{Belkacem} {et~al.}(2010){Belkacem}, {Dupret}, \&
  {Noels}}]{Belkacemetal2010}
{Belkacem}, K., {Dupret}, M.~A., \& {Noels}, A. 2010, \aap, 510, A6

\bibitem[{{Belkacem} {et~al.}(2009){Belkacem}, {Mathis}, {Goupil}, \&
  {Samadi}}]{Belkacemetal2009}
{Belkacem}, K., {Mathis}, S., {Goupil}, M.~J., \& {Samadi}, R. 2009, \aap, 508,
  345

\bibitem[{{Belkacem} {et~al.}(2008{\natexlab{a}}){Belkacem}, {Samadi}, \&
  {Goupil}}]{belkacem08}
{Belkacem}, K., {Samadi}, R., \& {Goupil}, M.~J. 2008{\natexlab{a}}, in Journal
  of Physics Conference Series, Vol. 118, Journal of Physics Conference Series,
  012028

\bibitem[{{Belkacem} {et~al.}(2008{\natexlab{b}}){Belkacem}, {Samadi},
  {Goupil}, \& {Dupret}}]{Belkacemetal2008}
{Belkacem}, K., {Samadi}, R., {Goupil}, M.-J., \& {Dupret}, M.-A.
  2008{\natexlab{b}}, \aap, 478, 163

\bibitem[{{B{\"o}hm-Vitense}(1958)}]{bohm1958}
{B{\"o}hm-Vitense}, E. 1958, \zap, 46, 108

\bibitem[{{Browning} {et~al.}(2004){Browning}, {Brun}, \&
  {Toomre}}]{Browningetal2004}
{Browning}, M.~K., {Brun}, A.~S., \& {Toomre}, J. 2004, \apj, 601, 512

\bibitem[{{Cantiello} {et~al.}(2009){Cantiello}, {Langer}, {Brott}, {de Koter},
  {Shore}, {Vink}, {Voegler}, {Lennon}, \& {Yoon}}]{Cantielloetal2009}
{Cantiello}, M., {Langer}, N., {Brott}, I., {et~al.} 2009, \aap, 499, 279

\bibitem[{{Deupree}(1990)}]{deupree90}
{Deupree}, R.~G. 1990, \apj, 357, 175

\bibitem[{{Deupree}(1995)}]{deupree95}
{Deupree}, R.~G. 1995, \apj, 439, 357

\bibitem[{{Dintrans} \& {Rieutord}(2000)}]{DR2000}
{Dintrans}, B. \& {Rieutord}, M. 2000, \aap, 354, 86

\bibitem[{{Edelmann} {et~al.}(2019){Edelmann}, {Ratnasingam}, {Pedersen},
  {Bowman}, {Prat}, \& {Rogers}}]{Edelmannetal2019}
{Edelmann}, P.~V.~F., {Ratnasingam}, R.~P., {Pedersen}, M.~G., {et~al.} 2019,
  \apj, 876, 4

\bibitem[{{Floquet} {et~al.}(2009){Floquet}, {Hubert}, {Huat}, {Fr{\'e}mat},
  {Janot-Pacheco}, {Guti{\'e}rrez-Soto}, {Neiner}, {de Batz}, {Leroy},
  {Poretti}, {Amado}, {Catala}, {Rainer}, {Diaz}, {Uytterhoeven}, {Andrade},
  {Diago}, {Emilio}, {Espinosa Lara}, {Fabregat}, {Martayan}, {Semaan}, \&
  {Suso}}]{floquet}
{Floquet}, M., {Hubert}, A.-M., {Huat}, A.-L., {et~al.} 2009, \aap, 506, 103

\bibitem[{{Fr{\'e}mat} {et~al.}(2005){Fr{\'e}mat}, {Zorec}, {Hubert}, \&
  {Floquet}}]{fremat}
{Fr{\'e}mat}, Y., {Zorec}, J., {Hubert}, A.-M., \& {Floquet}, M. 2005, \aap,
  440, 305

\bibitem[{{Gough}(1977)}]{Gough1977}
{Gough}, D.~O. 1977, \apj, 214, 196

\bibitem[{{Grimshaw}(1984)}]{grimshaw1984}
{Grimshaw}, R. 1984, Annual Review of Fluid Mechanics, 16, 11

\bibitem[{{Huat} {et~al.}(2009){Huat}, {Hubert}, {Baudin}, {Floquet}, {Neiner},
  {Fr{\'e}mat}, {Guti{\'e}rrez-Soto}, {Andrade}, {de Batz}, {Diago}, {Emilio},
  {Espinosa Lara}, {Fabregat}, {Janot-Pacheco}, {Leroy}, {Martayan}, {Semaan},
  {Suso}, {Auvergne}, {Catala}, {Michel}, \& {Samadi}}]{huat}
{Huat}, A.-L., {Hubert}, A.-M., {Baudin}, F., {et~al.} 2009, \aap, 506, 95

\bibitem[{{Kolmogorov}(1941)}]{Kolmogorov1941}
{Kolmogorov}, A. 1941, Akademiia Nauk SSSR Doklady, 30, 301

\bibitem[{{Lee}(2006)}]{lee2006}
{Lee}, U. 2006, \mnras, 365, 677

\bibitem[{{Lee}(2012)}]{lee2012}
{Lee}, U. 2012, \mnras, 420, 2387

\bibitem[{{Lee} \& {Baraffe}(1995)}]{lee_baraffe}
{Lee}, U. \& {Baraffe}, I. 1995, \aap, 301, 419

\bibitem[{{Lee} {et~al.}(2014){Lee}, {Neiner}, \& {Mathis}}]{lee2014}
{Lee}, U., {Neiner}, C., \& {Mathis}, S. 2014, \mnras, 443, 1515

\bibitem[{{Lee} \& {Saio}(1993{\natexlab{a}})}]{lee_saio}
{Lee}, U. \& {Saio}, H. 1993{\natexlab{a}}, \mnras, 261, 415

\bibitem[{{Lee} \& {Saio}(1993{\natexlab{b}})}]{leesaio1993}
{Lee}, U. \& {Saio}, H. 1993{\natexlab{b}}, \mnras, 261, 415

\bibitem[{{Lee} \& {Saio}(1997)}]{LeeSaio1997}
{Lee}, U. \& {Saio}, H. 1997, \apj, 491, 839

\bibitem[{{Mathis}(2009)}]{Mathis2009}
{Mathis}, S. 2009, \aap, 506, 811

\bibitem[{{Mathis} \& {de Brye}(2012)}]{mathis_debrye}
{Mathis}, S. \& {de Brye}, N. 2012, \aap, 540, A37

\bibitem[{{Mathis} {et~al.}(2014){Mathis}, {Neiner}, \& {Tran Minh}}]{MNT2014}
{Mathis}, S., {Neiner}, C., \& {Tran Minh}, N. 2014, \aap, 565, A47

\bibitem[{{Mathis} \& {Prat}(2019)}]{MP2019}
{Mathis}, S. \& {Prat}, V. 2019, \aap, 631, A26

\bibitem[{{Mathis} {et~al.}(2008){Mathis}, {Talon}, {Pantillon}, \&
  {Zahn}}]{mathis}
{Mathis}, S., {Talon}, S., {Pantillon}, F., \& {Zahn}, J. 2008, \solphys, 251,
  101

\bibitem[{{Mininni} \& {Pouquet}(2010)}]{mininni10}
{Mininni}, P.~D. \& {Pouquet}, A. 2010, Physics of Fluids, 22, 035105

\bibitem[{{Neiner} {et~al.}(2012{\natexlab{a}}){Neiner}, {Floquet}, {Samadi},
  {Espinosa Lara}, {Fr{\'e}mat}, {Mathis}, {Leroy}, {de Batz}, {Rainer},
  {Poretti}, {Mathias}, {Guarro Fl{\'o}}, {Buil}, {Ribeiro}, {Alecian},
  {Andrade}, {Briquet}, {Diago}, {Emilio}, {Fabregat}, {Guti{\'e}rrez-Soto},
  {Hubert}, {Janot-Pacheco}, {Martayan}, {Semaan}, {Suso}, \&
  {Zorec}}]{neinerStocha}
{Neiner}, C., {Floquet}, M., {Samadi}, R., {et~al.} 2012{\natexlab{a}}, \aap,
  546, A47

\bibitem[{{Neiner} {et~al.}(2009){Neiner}, {Guti{\'e}rrez-Soto}, {Baudin}, {de
  Batz}, {Fr{\'e}mat}, {Huat}, {Floquet}, {Hubert}, {Leroy}, {Diago},
  {Poretti}, {Carrier}, {Rainer}, {Catala}, {Thizy}, {Buil}, {Ribeiro},
  {Andrade}, {Emilio}, {Espinosa Lara}, {Fabregat}, {Janot-Pacheco},
  {Martayan}, {Semaan}, {Suso}, {Baglin}, {Michel}, \& {Samadi}}]{neiner2009}
{Neiner}, C., {Guti{\'e}rrez-Soto}, J., {Baudin}, F., {et~al.} 2009, \aap, 506,
  143

\bibitem[{{Neiner} {et~al.}(2012{\natexlab{b}}){Neiner}, {Mathis}, {Saio},
  {Lovekin}, {Eggenberger}, \& {Lee}}]{neinerMixing}
{Neiner}, C., {Mathis}, S., {Saio}, H., {et~al.} 2012{\natexlab{b}}, \aap, 539,
  A90

\bibitem[{{Pantillon} {et~al.}(2007){Pantillon}, {Talon}, \&
  {Charbonnel}}]{pantillon2007}
{Pantillon}, F.~P., {Talon}, S., \& {Charbonnel}, C. 2007, \aap, 474, 155

\bibitem[{{Pharasi} {et~al.}(2014){Pharasi}, {Kumar}, \&
  {Bhattacharjee}}]{pharasi14}
{Pharasi}, H.~K., {Kumar}, K., \& {Bhattacharjee}, J.~K. 2014, \pre, 89, 023009

\bibitem[{{Pin{\c c}on} {et~al.}(2016){Pin{\c c}on}, {Belkacem}, \&
  {Goupil}}]{pincon2016}
{Pin{\c c}on}, C., {Belkacem}, K., \& {Goupil}, M.~J. 2016, \aap, 588, A122

\bibitem[{{Rieutord}(2009)}]{rieutord2009}
{Rieutord}, M. 2009, in Lecture Notes in Physics, Berlin Springer Verlag, Vol.
  765, The Rotation of Sun and Stars, ed. J.-P. {Rozelot} \& C.~{Neiner},
  101--121

\bibitem[{{Rivinius} {et~al.}(2003){Rivinius}, {Baade}, \&
  {{\v{S}}tefl}}]{rivinius2003}
{Rivinius}, T., {Baade}, D., \& {{\v{S}}tefl}, S. 2003, \aap, 411, 229

\bibitem[{{Rivinius} {et~al.}(2013){Rivinius}, {Carciofi}, \&
  {Martayan}}]{rivinius2013}
{Rivinius}, T., {Carciofi}, A.~C., \& {Martayan}, C. 2013, \aapr, 21, 69

\bibitem[{{Rogers}(2015)}]{rogers2015}
{Rogers}, T.~M. 2015, \apjl, 815, L30

\bibitem[{{Rogers} {et~al.}(2013){Rogers}, {Lin}, {McElwaine}, \&
  {Lau}}]{rogers2013}
{Rogers}, T.~M., {Lin}, D.~N.~C., {McElwaine}, J.~N., \& {Lau}, H.~H.~B. 2013,
  \apj, 772, 21

\bibitem[{{Saio}(2018)}]{saio2018arXiv}
{Saio}, H. 2018, arXiv e-prints [\eprint[arXiv]{1812.01253}]

\bibitem[{{Saio} {et~al.}(2018){Saio}, {Kurtz}, {Murphy}, {Antoci}, \&
  {Lee}}]{saio2018MNRAS}
{Saio}, H., {Kurtz}, D.~W., {Murphy}, S.~J., {Antoci}, V.~L., \& {Lee}, U.
  2018, \mnras, 474, 2774

\bibitem[{{Samadi} {et~al.}(2010){Samadi}, {Belkacem}, {Goupil}, {Dupret},
  {Brun}, \& {Noels}}]{Samadietal2010}
{Samadi}, R., {Belkacem}, K., {Goupil}, M.~J., {et~al.} 2010, \apss, 328, 253

\bibitem[{{Samadi} \& {Goupil}(2001)}]{Samadietal2001}
{Samadi}, R. \& {Goupil}, M.-J. 2001, \aap, 370, 136

\bibitem[{{Samadi} {et~al.}(2003){Samadi}, {Nordlund}, {Stein}, {Goupil}, \&
  {Roxburgh}}]{Samadietal2003}
{Samadi}, R., {Nordlund}, {\AA}., {Stein}, R.~F., {Goupil}, M.~J., \&
  {Roxburgh}, I. 2003, \aap, 404, 1129

\bibitem[{{Schatzman}(1993)}]{schatzman1993}
{Schatzman}, E. 1993, \aap, 279, 431

\bibitem[{{Schenk} {et~al.}(2002){Schenk}, {Arras}, {Flanagan}, {Teukolsky}, \&
  {Wasserman}}]{Schenketal2002}
{Schenk}, A.~K., {Arras}, P., {Flanagan}, {\'E}.~{\'E}., {Teukolsky}, S.~A., \&
  {Wasserman}, I. 2002, \prd, 65, 024001

\bibitem[{{Stein}(1967)}]{Stein1967}
{Stein}, R.~F. 1967, \solphys, 2, 385

\bibitem[{{Townsend}(1965)}]{townsend1965}
{Townsend}, A.~A. 1965, Journal of Fluid Mechanics, 22, 241

\bibitem[{{Vadas} \& {Fritts}(2001)}]{vadas2001}
{Vadas}, S.~L. \& {Fritts}, D.~C. 2001, Journal of Atmospheric Sciences, 58,
  2249

\bibitem[{{Walker} {et~al.}(2005){Walker}, {Kuschnig}, {Matthews}, {Cameron},
  {Saio}, {Lee}, {Kambe}, {Masuda}, {Guenther}, {Moffat}, {Rucinski},
  {Sasselov}, \& {Weiss}}]{walker}
{Walker}, G.~A.~H., {Kuschnig}, R., {Matthews}, J.~M., {et~al.} 2005, \apjl,
  635, L77

\end{thebibliography}

\begin{appendix}

\section{Formalism for stochastically excited g modes modified by rapid rotation}\label{appendix}

In this appendix, we describe the equations derived and implemented in the Tohoku code for the treatment of stochastically excited gravity modes modified by rapid rotation. 

\subsection{Stochastic excitation}

The inhomogeneous wave equation taking the Coriolis
acceleration into account and assuming a solid body rotation can be written

\begin{eqnarray}
\label{wave_equation}
\left(\derivp{^2}{t^2} - \vec L_\Omega \right) \vec v_{\rm osc} + {\vec  {\cal C}_{\rm osc}} = \vec {\cal S}_t\,.
\end{eqnarray}

All physical quantities are split into an equilibrium term and a perturbation.
In the following equations, the subscripts 1 and 0 denote Eulerian perturbations
and equilibrium quantities respectively, except for the velocity where the
subscript 1 has been dropped to ease the notation. The latter is split into two
contributions, namely the oscillation velocity ($\vec v_{\rm osc}$) and the
turbulent velocity field ($\vec u_t$), such that $\vec u= \vec v_{\rm osc} +
\vec u_t$.  $\vec L_\Omega$ is the linear oscillation operator which, in
presence of uniform rotation, becomes

\begin{eqnarray}
\lefteqn{\vec L_\Omega = \vec \nabla \left[ \alpha_s \vec v_{\rm osc} \, \cdot \, \vec \nabla s_0 + c_s^2 \vec \nabla \left( \rho_0 \vec v_{\rm osc} \right) \right] 
- \vec g_{\rm eff}  \nabla \cdot \left( \rho_0 \vec v_{\rm osc} \right)} \nonumber \\
& &- \rho_0 \Omega \frac{\partial^2 \vec v_{\rm osc}}{\partial t \partial \varphi} 
- 2 \, \rho_0 \vec \Omega \times \derivp{\vec v_{\rm osc}}{t}\,,
\end{eqnarray}

\noindent where $c_s$ is the sound velocity and $\alpha_{s}=\left(\partial
p/\partial s\right)_{\rho}$. Next, the operator 

\begin{eqnarray}
\mathcal{C}_{\rm osc} &=& \derivp{}{t} \Big[ \derivp{\left(\rho_1 \vec v_{\rm osc}\right)}{t} + 2\, \vec \nabla : \left( \rho_0 \vec v_{\rm osc} \vec u_t \right) 
 \nonumber \\
 &+&  \rho_1 \Omega  \derivp{\vec v_{\rm osc}}{\varphi} + 2 \, \rho_1 \vec \Omega \times \vec v_{\rm osc} \nonumber \\
 &+& \vec \nabla \left(\alpha_s \vec v_{\rm osc} \cdot \vec \nabla s_1 
 + c_s^2 \vec \nabla \cdot (\rho_1 \vec v_{\rm osc}) \right) \Big]\,
\end{eqnarray}

\noindent involves both turbulent and pulsational velocities and contributes to
the linear dynamical damping \citep[see][for details]{Samadietal2001}. Finally,
the $\mathcal{S}_t$ operator that contains the source terms of the inhomogeneous
wave equation (\eq{wave_equation}) is given by

\begin{eqnarray}
\label{Source}
\mathcal{S}_t &= \displaystyle{- \derivp{}{t}} \vec \nabla : (\rho_0 \vec u_t \vec u_t) + \vec \nabla (\alpha_s \vec u_t \cdot \vec \nabla s_1) + 
 \mathcal{S}_\Omega +  \mathcal{S}_M
\end{eqnarray}

\noindent with
\begin{eqnarray}
\label{termes_sources}
\mathcal{S}_\Omega &=&  - \derivp{}{t}\left[ \rho_1 \left( \Omega \derivp{}{\varphi} \vec u_t  
 - 2 \vec \Omega \times \vec u_t 
\right)\right], \\
 \mathcal{S}_M &=& \derivp{}{t} (\rho_1 {\bf g}_1) + \vec \nabla \left[c_s^2 \vec \nabla \cdot (\rho_1 \vec u_t)\right] 
 - \vec g_{\rm eff} \cdot \vec \nabla (\rho_1 \vec u_t) \nonumber \\
 &-& \derivp{^2}{t^2} (\rho_1 \vec u_t) + \mathcal{L}_t,
\end{eqnarray}

\noindent where $\mathcal{L}_t$ contains the linear terms source\footnote{Linear
terms are defined as the product of an equilibrium quantity and a fluctuating
one.}. The first two terms of \eq{Source} correspond to the Reynolds stress and
entropy contributions, respectively. The rotational contributions are the two
following terms in \eq{Source}, that are respectively related to the Doppler
effect and the Coriolis acceleration. Finally, the next four terms are neglected
since they scale as $\mathcal{M}^3$ where $\mathcal{M}$ is the Mach number
\citep[see][for details]{Samadietal2001}. Therefore, those terms are negligible
in comparison to the Reynolds stress contribution that scales as
$\mathcal{M}^2$. Note that in the case of "free" turbulence where the oscillations do not have any effect on the turbulent flow, we have $\rho_1\approx\rho_t$.

\subsection{Mean square amplitude}

Using \eq{wave_equation}, we can determine the mean square amplitude of modes
excited stochastically. 

First, we recall that in a rapidly rotating star, a single $l$ value does no
longer represent a pulsation mode for a given $m$ value because of the
latitudinal couplings between the different spherical harmonics owing to the
Coriolis and the centrifugal accelerations. For this reason the eigenfunctions
for a fixed $m$ are expanded into a series of terms proportional to spherical
harmonics $Y_{l,m}\left(\theta,\varphi\right)$ with $l\ge\vert m\vert$. We
express the spatial dependence of the displacement vector $\vec{\xi}$ 
as

\begin{equation}
\vec{\xi} = \displaystyle\sum_{j=1}^{J}\left\{\xi_{r;m}^{l_j}Y^{m}_{l_j}{\hat{\bf e}}_{r} + \xi_{H;m}^{l_j}{\vec\nabla}_{H}Y^{m}_{l_j} +\xi_{T;m}^{l_{j^{\,'}}}\left[{\vec\nabla}_{H}Y^m_{l_{j^{\,'}}} \times{\hat{\bf e}}_r\right]\right\},
\label{displacement}
\end{equation}


\noindent where ${\vec\nabla}_{H} = \displaystyle{\frac{\partial}{\partial\theta}}{\hat{\bf
e}}_{\theta} + \displaystyle{\frac{1}{\sin \theta}\frac{\partial}{\partial\varphi}}{\hat{\bf
e}}_{\varphi}$ and $\left(r,\theta,\varphi\right)$ are the usual spherical coordinates, $l_j=\vert m\vert+2\left(j-1\right)+I$ and $l_{j^{\,'}}=l_{j}+1-2I$
with $I=0$ for even modes and $I=1$ for odd ones. The terms proportional to
$\xi_{T;m}^{l_{j^{\,}}}$ represent a toroidal displacement needed because of the Coriolis
acceleration term in the momentum equation.

In addition, the wave velocity field ($\vec v$) is related to the displacement
($\vec \xi$) by 

\begin{equation}
\label{v_nodiff}
\vec v = A\,i(\omega_0+m\Omega)\,\vec \xi\,e^{i \omega_0 t} \, ,
\end{equation} 

\noindent where $\omega_0$ is the eigenfrequency and $A(t)$ is the amplitude due
to the turbulent forcing.

Following \cite{Belkacemetal2009} and generalising the results to the case of
rapid rotation, one then obtains the mean square amplitude

\begin{equation}
\label{mean_square_amplitude_trapping}
<|A|^2> = \frac{1}{8 \eta \omega_{0}^{2}}\left|\frac{1}{b}\right|^{2}\left( C_R^2 + C_S^2 + C_\Omega^2 + C_{c}^{2}  \right),
\end{equation}

\noindent where the operator $<\,.\,.\,.>$ denotes a statistical average
performed on an infinite number of independent realisations, $\eta$ is the
damping rate computed by the non-adiabatic oscillation code in the radiative
flux approximation (i.e. $\delta\left(\vec\nabla\cdot{\vec F}_{\rm
conv}\right)=0$ where ${\vec F}_{\rm conv}$ is the convective flux; see e.g.
\cite{lee2006}) as expected in massive stars \citep[see the discussion
in][]{Belkacemetal2010,Samadietal2010}. 

We introduce the modified mode inertia in the case of rapid rotation
\citep[][]{Schenketal2002}:

\begin{equation}
b=\sum_{j}\int_{0}^{R_{\rm CZ}}I_{b}^{j}\left(r\right)\rho_{0}\left(r\right)r^2{\rm d}r,
\label{normalisation}
\end{equation}

\noindent where
\begin{eqnarray}
I_{b}^{j}\left(r\right)&=&\left(\xi_{r;m}^{l_j}\right)^{*}\left[\xi_{r;m}^{l_j}-\frac{i\Omega}{\omega_{0}}\left(C_{m}^{l_j}+D_{m}^{l_j}\right)\right]\nonumber\\
&&+{L_{j}}^{2}\left(\xi_{H;m}^{l_j}\right)^{*}\left[\xi_{H;m}^{l_j}-\frac{i\Omega}{\omega_{0}}\left(E_{m}^{l_j}+F_{m}^{l_j}\right)\right]\nonumber\\
&&+{L_{j^{\,'}}}^{2}\left(\xi_{T;m}^{l_j^{\,'}}\right)^{*}\left[\xi_{T;m}^{l_j^{\,'}}-\frac{i\Omega}{\omega_{0}}\left(G_{m}^{l_j^{\,'}}+H_{m}^{l_j^{\,'}}\right)\right]
\end{eqnarray}

\noindent with
\begin{eqnarray}
C_{m}^{l_j}&=&\frac{1}{{L_{j}}^{2}}\left[l_jB_m^{l_j}\xi_{T;m}^{l_j-1}-\left(l_j+1\right)B_{m}^{l_j+1}\xi_{T;m}^{l_j+1}\right]\nonumber\\
D_{m}^{l_j}&=&-im\xi_{H;m}^{l_j}\nonumber\\
E_{m}^{l_j}&=&\frac{1}{{L_j}^{2}}\left[B_{l_j}^{m}\xi_{T;m}^{l_j-1}+B_{m}^{l_j+1}\xi_{T;m}^{l_j+1}\right]\nonumber\\
F_{m}^{l_j}&=&-\frac{im}{{L_{j}}^{2}}\left[\xi_{r;m}^{l_j}+\xi_{H;m}^{l_j}\right]\nonumber\\
G_{m}^{l_j^{\,'}}&=&-\frac{im}{{L_{j^{\,'}}}^{2}}\xi_{T;m}^{l_{j^{\,'}}}\nonumber\\
H_{m}^{l_{j^{\,'}}}&=&l_{j^{\,'}}A_{m}^{l_{j^{\,'}}}\left[\xi_{r;m}^{l_{j^{\,'}}-1}-\left(l_{j^{\,'}}-1\right)\xi_{H;m}^{l_{j^{\,'}}-1}\right]\nonumber\\
&&-\left(l_{j^{\,'}}+1\right)A_{m}^{l_{j^{\,'}}+1}\left[\xi_{r;m}^{l_{j^{\,'}}+1}+\left(l_{j^{\,'}}+2\right)\xi_{H;m}^{l_{j^{\,'}}+1}\right]\nonumber
\end{eqnarray}

\noindent where
\begin{equation*}
A_{m}^{l}=\frac{1}{l^2}\sqrt{\frac{l^2-m^2}{4l^2-1}}\quad\hbox{and}\quad B_{m}^{l}=l^2\left(l^2-1\right)A_{m}^{l}
\end{equation*}

\noindent and $L_{l}^{2}=l\left(l+1\right)$. 

Finally, $C_R^2$ is the Reynolds stress contribution, $C_S^2$ is the one related
to entropy fluctuations, $C_\Omega^2$ contains the contributions of the Coriolis
acceleration and the Doppler term, and $C_{c}$ represents the cross-source
terms, that is the interferences between the various excitation sources.

\subsubsection{Reynolds stress contribution}

We follow \cite{Belkacemetal2009} but have modified their R3 term to account for contributions that are important for rapid rotation. As a result, the turbulent Reynolds stress contribution is:

\begin{eqnarray}
\label{C2R_ref}
C_R^2 & =  &  4\pi^{3}   \int_{{\mathcal V}_{\rm CZ}}  \textrm{d}m   \,  \;\sum_{j}R_{m}^{j}(r)~ S_R(\omega_0)\; ,
 \end{eqnarray}
 
\noindent where ${\mathcal V}_{\rm CZ}$ is the volume of the convective regions,
with
\begin{equation}
R_{m}^{j}(r)=R_{1;m}^{j}(r)-R_{2;m}^{j}(r)+R_{3;m}^{j}(r)
\end{equation}
(we refer the reader to Eq. A.8 in \cite{Belkacemetal2009}). We have:
\begin{eqnarray}
\label{gammabeta}
R_{1;m}^{j}(r) &=&   2 ~  \left| \deriv{\xi_{r;m}^{l_j}}{r}  \right|^2  +  4 ~    \left| \frac{\xi_{r;m}^{l_j}}{r}  \right|^2\nonumber\\
&+&  ~ L_{l_j}^{2} \left| \mathcal{A}_{R;m}^{l_j}  \right|^2 - 2 L_{l_j}^{2} \left[\frac{\left(\xi_{r;m}^{l_j}\right)^{*} \xi_{H,m}^{l_j}}{r^2} +{\rm c.c.}\right] 
\nonumber \\
&+& 2 L_{l_j}^{2}\left(L_{l_j}^{2}-1\right) \left| \frac{\xi_{H;m}^{l_j}}{r}  \right|^2 
\nonumber \\
&+&  L_{l_j^{\,'}}^{2}\left(L_{l_j^{\,'}}^{2}-2\right) \left| \frac{\xi_{T;m}^{l_{j^{\,'}}}}{r}  \right|^2  
\nonumber \\
 &+&L_{l_j^{\,'}}^{2}\left| \mathcal{B}_{R;m}^{l_{j^{\,'}}}  \right|^2\,,
\end{eqnarray}

\noindent where 
\begin{eqnarray}
\mathcal{A}_{R;m}^{l_j} &=& \deriv{\xi_{H;m}^{l_j}}{r} +\frac{1}{r}\left(\xi_{r;m}^{l_j}-\xi_{H;m}^{l_j}\right)\nonumber\\
\mathcal{B}_{R;m}^{l_j^{\,'}} &=&  \deriv{\xi_{T;m}^{l_j^{\,'}}}{r} - \frac{\xi_{T;m}^{l_j^{\,'}}}{r}\,,
\end{eqnarray}

\noindent while c.c denotes the complex conjugate. This allows us to compute
\begin{equation}
R_{2;m}^{j}(r)=2\alpha R_{1;m}^{j}(r)  
\end{equation}
with $\alpha=1/3$. Finally, we have to compute
\begin{equation}
R_{3;m}^{j}(r)=\beta \left(R_{1;m}^{j}(r)+2\int_{4\pi}\frac{{\rm d}{\widetilde\Omega}}{4\pi}\sum_{i,j;i\not=j}\left(B_{ii}^*B_{jj}+B_{ii}B_{jj}^*\right)\right),     
\end{equation}
where we refer the reader to the formalism explained in the appendix A of \cite{Belkacemetal2009} for the definition of the $B_{ii}$ coefficients (see their Equation A.17) while $\beta=1/5$, ${\widetilde \Omega}$ is the solid angle with ${\rm d}{\widetilde\Omega}=\sin\theta\,{\rm d}\theta\,{\rm d}\varphi$ and $^{*}$ is the complex conjugate. We have: 
\begin{eqnarray}
\lefteqn{\sum_{i,j;i\not=j}\left(B_{ii}^*B_{jj}+B_{ii}B_{jj}^*\right)=2\left|\frac{\xi_r}{r}\right|^2+2\left(\frac{\partial\xi_r^*}{ \partial r}\frac{\xi_r}{r}+
\frac{\partial \xi_r}{\partial r}\frac{\xi_r^*}{r} \right)} \nonumber\\
&&+\left[\left(\frac{\partial\xi_r^*}{\partial r}+\frac{\xi_r^*}{r}\right)\nabla_H\cdot\pmb{\xi}_H
+\left(\frac{\partial\xi_r}{\partial r}+\frac{\xi_r}{r}\right)\nabla_H\cdot\pmb{\xi}^*_H\right]\nonumber\\
&&+\frac{1}{r}\frac{\partial\xi_\theta}{\partial\theta}\left(\frac{\xi_\theta}{r}\frac{\cos\theta}{\sin\theta}+\frac{1}{r\sin\theta}\frac{\partial\xi_\phi}{\partial\phi}\right)^*+{\rm c.c.}\, ,\nonumber
\end{eqnarray}
where ${\rm c.c.}$ means the complex conjugate.

Upon integration over the surface, we obtain
\begin{eqnarray}
\lefteqn{\int_{4\pi} \frac{{\rm d}{\widetilde \Omega}}{4\pi} \sum_{i,j;i\not=j}\left(B_{ii}^*B_{jj}+B_{ii}B_{jj}^*\right)=\nonumber}\\
&&\sum_{j}\left\{ 2\left|\frac{\xi_{r;m}^{l_j}}{r} \right|^2
+2\left(\frac{{\rm d}\left(\xi_{r;m}^{l_j}\right)^*}{{\rm d}r}\frac{\xi_{r;m}^{l_j}}{r}+{\rm c.c.}
\right)\right.\nonumber\\
&&{\left.-L_{l_j}^2\left[\left(\frac{{\rm d}\left(\xi_{r;m}^{l_j}\right)^*}{{\rm d}r}+\frac{\left(\xi_{r;m}^{l_j}\right)^*}{r}\right)\frac{\xi_{H;m}^{l_j}}{r}+{\rm c.c.}
\right]\right\}}\nonumber\\
&&+\int_{4\pi}\frac{{\rm d}{\widetilde \Omega}}{4\pi}\left[\frac{1}{r}\frac{\partial\xi_\theta}{ \partial\theta}\left(\frac{\xi_\theta}{r}\frac{\cos\theta}{\sin\theta}+\frac{1}{r\sin\theta}\frac{\partial\xi_\varphi}{\partial\varphi}\right)^*+{\rm c.c.}\right].
\end{eqnarray}
To evaluate the last terms, we use the following procedure:
\begin{eqnarray}
\frac{1}{r}\frac{\partial\xi_\theta}{\partial\theta}&\!\!=\!\!&\sum_j\left\{\frac{\xi_{H;m}^{l_j}}{r}\left[-\Lambda_{l_j}Y_{l_j}^m-\frac{1}{\sin^2\theta}\left(\cos\theta\sin\theta\frac{\partial Y_{l_j}^m}{\partial\theta}-m^2Y_{l_j}^m\right)\right]\right.\nonumber\\
&&{\left.+ m\left(\rm i \frac{\xi_{T;m}^{l_j^{\,'}}}{r} \right)\frac{1}{\sin^2\theta}\left(\sin\theta\frac{\partial Y_{l_j^{\,'}}^m}{\partial\theta}-\cos\theta Y_{l_j^{\,'}}^m\right)\right\}}\nonumber\\
&=&\sum_{j,k}\left[\frac{\xi_{H;m}^{l_j}}{r}\left(-\Lambda_{l_jl_k}+A_{l_jl_k}\right)+m\left(\rm i \frac{\xi_{T;m}^{l_j^{\,'}}}{r} \right)B_{l_j^{\,'}l_k}\right]Y_{l_k}^m\,,
\end{eqnarray}
\begin{eqnarray}
\lefteqn{\frac{1}{r\sin\theta}\frac{\partial\xi_\varphi}{\partial \varphi}+\frac{\cos\theta}{r\sin\theta}\xi_\theta=\sum_j\left\{\frac{\xi_{H;m}^{l_j}}{r}\frac{1}{\sin^2\theta}
\left(\cos\theta\sin\theta\frac{\partial Y_{l_j}^m}{\partial\theta}-m^2Y_{l_j}^m\right)\right.}\nonumber\\
&&{\left.+m\left(\rm i \frac{\xi_{T;m}^{l_j^{\,'}}}{r} \right)\frac{1}{\sin^2\theta}\left(\cos\theta Y_{l_j^{\,'}}^m-\sin\theta\frac{\partial Y_{l_j^{\,'}}^m}{\partial \theta}\right)\right\}}\nonumber\\
&=&-\sum_{j,k}\left[\frac{\xi_{H;m}^{l_j}}{r}A_{l_j l_k}+m\left(\rm i \frac{\xi_{T;m}^{l_j^{\,'}}}{r}\right)B_{l_j^{\,'}l_{k}}\right]Y_{l_k}^m
\end{eqnarray}
where we introduce the matrices $(\Lambda_{lk})$, $(A_{lk})$, and $(B_{lk})$ defined by
\be
\Lambda_{l_j l_k}=\delta_{l_j l_k}L_{l_j}^2\, ,
\ee
\be
A_{l_j l_k}=-\int_{4\pi} \frac{d{\widetilde\Omega}}{4\pi}\frac{1}{\sin^2\theta}\left(\cos\theta\sin\theta\frac{\partial Y_{l_j}^m}{\partial\theta}-m^2Y_{l_j}^m\right)(Y_{l_k}^m)^*\, ,
\ee
\be
B_{l_j^{\,'} l_k}=\int_{4\pi} \frac{d{\widetilde\Omega}}{4\pi} \frac{1}{\sin^2\theta}\left(\sin\theta\frac{\partial Y_{l_j^{\,'}}^m}{\partial\theta}-\cos\theta Y_{l_j^{\,'}}^m\right)(Y_{l_k}^m)^*\, .
\ee
This leads to:
\begin{eqnarray}
\lefteqn{\int_{4\pi} \frac{d{\widetilde\Omega}}{4\pi}\left[\frac{1}{r}\frac{\partial\xi_\theta}{\partial\theta}\left(\frac{\xi_\theta}{r}\frac{\cos\theta}{\sin\theta}+\frac{1}{r\sin\theta}\frac{\partial\xi_\varphi}{\partial\varphi}\right)^*+{\rm c.c.}\right]}\nonumber\\
&=&-\sum_{j,k}\left(\frac{\xi_{H;m}^{l_j}}{r}\left(\frac{\xi_{H;m}^{l_k}}{r}\right)^*+{\rm c.c.}\right)\left(-A_{l_j l_k}\Lambda_{l_k}+\sum_n A_{l_j l_n}A_{l_k l_n}\right)\nonumber\\
&&-\sum_{j,k}\left[i\frac{\xi_{T;m}^{l_j^{\,'}}}{r}\left(\frac{\xi_{H;m}^{l_k}}{r}\right)^*+{\rm c.c.}\right]\left(-B_{l_j^{\,'}l_k}\Lambda_{l_k}+2\sum_n B_{l_j^{\,'}l_n}A_{l_kl_n}\right)\nonumber\\
&&-\sum_{j,k}\left[i\frac{\xi_{T;m}^{l_j^{\,'}}}{r}\left(i\frac{\xi_{T;m}^{l_k^{\,'}}}{r}\right)^*+{\rm c.c.}\right]\sum_n B_{l_j^{\,'}l_n}B_{l_k^{\,'}n}.
\end{eqnarray}
We finally obtain:
\begin{eqnarray}
R_{m}^{j}(r) &=&   \frac{16}{15} ~  \left| \deriv{\xi_{r;m}^{l_j}}{r}  \right|^2  +  \frac{44}{15} ~    \left| \frac{\xi_{r;m}^{l_j}}{r}  \right|^2 + \frac{4}{5}\left(\frac{\xi_{r;m}^{l_j}}{r}\frac{{\rm d}\left(\xi_{r;m}^{l_j}\right)^*}{{\rm d}r}+{\rm c.c.}\right)\nonumber\\
&+& \frac{8}{5} ~ L_{l_j}^{2} \left| \mathcal{A}_{R;m}^{l_j}  \right|^2 - \frac{22}{15} L_{l_j}^{2} \left[\frac{\left(\xi_{r;m}^{l_j}\right)^{*} \xi_{H,m}^{l_j}}{r^2} +{\rm c.c.}\right]
\nonumber \\
&-& \frac{2}{5}~ L_{l_j}^2\left[\left(\frac{{\rm d}\left(\xi_{r;m}^{l_j}\right)^*}{{\rm d}r}\right)\frac{\xi_{H;m}^{l_j}}{r}+{\rm c.c.}
\right]\nonumber\\
&+& \frac{16}{15} L_{l_j}^{2}\left(L_{l_j}^{2}-1\right) \left| \frac{\xi_{H;m}^{l_j}}{r}  \right|^2 
\nonumber \\
&-&\frac{2}{5}\sum_{j,k}\left(\frac{\xi_{H;m}^{l_j}}{r}\left(\frac{\xi_{H;m}^{l_k}}{r}\right)^*+{\rm c.c.}\right)\left(-A_{l_j l_k}\Lambda_{l_k}+\sum_n A_{l_j l_n}A_{l_k l_n}\right)\nonumber\\
&+& \frac{8}{5} ~ L_{l_j^{\,'}}^{2}\left(L_{l_j^{\,'}}^{2}-2\right) \left| \frac{\xi_{T;m}^{l_{j^{\,'}}}}{r}  \right|^2 
\nonumber \\
&+& \frac{8}{5} ~ L_{l_j^{\,'}}^{2}\left| \mathcal{B}_{R;m}^{l_{j^{\,'}}}  \right|^2\ \nonumber\\
&-&\frac{2}{5}\sum_{j,k}\left[i\frac{\xi_{T;m}^{l_j^{\,'}}}{r}\left(i\frac{\xi_{T;m}^{l_k^{\,'}}}{r}\right)^*+{\rm c.c.}\right]\sum_n B_{l_j^{\,'}l_n}B_{l_k^{\,'}n}\nonumber\\
&-&\frac{2}{5}\sum_{j,k}\left[i\frac{\xi_{T;m}^{l_j^{\,'}}}{r}\left(\frac{\xi_{H;m}^{l_k}}{r}\right)^*+{\rm c.c.}\right]\left(-B_{l_j^{\,'}l_k}\Lambda_{l_k}+2\sum_nB_{l_j^{\,'}l_n}A_{l_kl_n}\right)\,.\nonumber\\
\end{eqnarray}

Furthermore,

\begin{equation}
S_R(\omega_0) = \,\int_{0}^{\infty}  \frac {\textrm{d}k} {k^2 }~E^2(k) ~\int_{-\infty}^{\infty} \textrm{d}\omega 
~\chi_k( \omega + \omega_0) ~\chi_k( \omega )\, ,
\label{fct_source}
 \end{equation}
 
\noindent where $(k,\omega)$ are the wavenumber and frequency of the turbulent
eddies and $E(k, \omega)$ is the turbulent kinetic energy spectrum, which is
expressed as the product $E(k) \, \chi_k(\omega)$  for isotropic turbulence. Indeed, as it will be discussed in detail in \S A.2.3., here we follow \cite{Samadietal2010} and \cite{Samadietal2001} by making the quasi-normal approximation and the assumptions of stationary, incompressible, homogeneous, and isotropic turbulence. These hypotheses allow us to explicitly relate the Reynolds stress integrals to the energy spectra. We refer the reader to Eqs. 25, 26, 27, 28 and 34 in \cite{Samadietal2001} for more technical details. The same assumptions are made to express the entropy fluctuations contribution. 
$E\left(k\right)$ is the spatial
kinetic energy spectrum, while $\chi_{k}\left(\omega\right)$ is the eddy-time
correlation function. In Sect.~\ref{turb}, a detailed discussion of how to
describe those functions will be given.

\subsubsection{Entropy fluctuations contribution}

As shown by \cite{Samadietal2001} and discussed briefly in
\cite{Samadietal2010}, the Reynolds stress contribution is not the unique source
of excitation. One also has to evaluate the contribution to the excitation by
the entropy fluctuations advection. Following \cite{Belkacemetal2008}, this
entropy source term depends on the mode compressibility. Consequently, since the
divergence of the toroidal component, which is the rotational part of the
spherical harmonic, vanishes, one can consider the poloidal contribution only.
Therefore, the final expression for the contribution of entropy fluctuations remains
the same for rapidly rotating stars as in non-rotating stars 
\citep{Belkacemetal2008}, that is

\eqn{
\label{C_S_ref}
C_S^2  = \frac{4 \pi^3 \, \mathcal{H}}{\omega_0^2}  \, 
\int_{{\mathcal V}_{\rm CZ}} \textrm{d}^3 x_0 \, \alpha_s^2 \, \sum_{j}\left ( {\mathcal A}_{S;m}^{l_j} + {\mathcal B}_{S;m}^{l_j} \right ) \,
\mathcal{S}_S(\omega_0) \, ,
}

\noindent where $\mathcal{H}$ is the anisotropy factor introduced in
\cite{Samadietal2001} which, in the current assumption (isotropic turbulence), 
is equal to $4/3$. In addition,

\eqna{
\label{C_S_ref2}
{\mathcal A}_{S;m}^{l_j} & \equiv & \frac{1 }{r^2}  {\mathcal D}_{m}^{l_j}\,  \left|   \, 
\deriv{\left( \ln \mid \alpha_s \mid \right)}{\ln r}
 - \deriv{{\mathcal D}_{m}^{l_j}}{\ln r}  \right|^2\,,\label{eqn:Al}
\\
\label{Bell}
{\mathcal B}_{S;m}^{l_j} & \equiv &  \frac{1 }{r^2} \, {L_j}^2 \, \left| {\mathcal D}_{m}^{l_j}\right| ^2\,,
\label{eqn:Bl}}

\noindent where
\begin{equation}\label{C_S_ref3}
{\mathcal D}_{m}^{l_j}\left(r,l_j\right) \equiv  {\mathcal D}_r - \frac{{L_j}^2} {r} \, \xi_{H;m}^{l_j} \; , \;
{\mathcal D}_r \equiv \frac{1}{r^2} \, \deriv{}{r} \left (  r^2  \xi_{r;m}^{l_j} \right )
\end{equation}

\noindent and
\begin{equation}
\mathcal{S}_S(\omega_0)  \equiv  \int_{0}^{\infty} \frac{\textrm{d}k}{k^4}\,E(k)
\, E_s(k) \, \int_{-\infty}^{\infty} \textrm{d}\omega \,
\chi_k(\omega_0+\omega)\,  \chi_k(\omega)\,,\label{eqn:FS}
\end{equation}

\noindent where $E_{s}\left(k\right)$ is the spatial entropy spectrum while
$\chi_{k}\left(\omega\right)$ is the same eddy-time correlation function as in
the Reynolds contribution.

\subsubsection{Rotational contributions}

To get clues about the contributions of the rotational terms, we express them in
a suitable form following \cite{Belkacemetal2009}:

\begin{eqnarray}
\lefteqn{\derivp{}{t} \left( 2 \, \rho_1 \vec \Omega \times  \vec u_t \right)-\derivp{}{t} \left( \rho_1 \Omega \derivp{\vec u_t }{\varphi} \right)=}\nonumber\\  
&&-\left[2 \vec \Omega \times \vec u_t +\Omega \derivp{\vec u_t}{\varphi} \right]\vec \nabla \cdot \left( \vec \rho_0 \vec u_t \right)\,.
\label{rotation}
\end{eqnarray}

Those terms vanish under the assumption that the turbulent field is anelastic.
This approximation is verified because we are dealing with excitation of modes
in convective zones. Indeed, the energy equation at the first order, together
with the mass conservation equation (i.e. $\vec \nabla \cdot (\rho_0 \vec u_t) =
0$), is equivalent to a quasi-isentropic equilibrium state.

In addition, by using dimensional analysis \citep{Samadietal2001}, it appears
that all those terms scale as the Mach number to the third ($\mathcal{M}^3$)
because they present a dependency to the perturbed mass flux in $\rho_1 \, \vec
u_t$. Compared to the Reynolds contribution, which scales as $\mathcal{M}^2$,
all rotational contributions thus become negligible in the subsonic regime ($\mathcal{M}<1$). 

\subsection{Turbulence modelling}\label{turb}

To model turbulence, we first have to specify the spatial kinetic energy
spectrum ($E\left(k\right)$) and the spatial energy spectrum for entropy
($E_{s}\left(k\right)$). 

While it is tempting to adopt relatively simple models for the nature of rotating convection, the stochastic excitation mechanisms that can drive gravito-inertial waves are subtle as they depend upon the precise nature of the turbulent convection in the driving region. 
The most widely used approach is to assume that the convective motions driving the waves follow a Heisenberg-Kolmogorov spectrum, which is also the tack taken here.  In contrast to the well-tested Heisenberg-Kolmogorov theory, the spectral form of the Reynolds stress correlations that arise in buoyancy-driven turbulent rotating convection are highly model dependent and as yet do not have a closed form solution.
This difficulty arises because of the broken symmetries of inhomogeneous and anisotropic turbulence that exist in these systems. Given that the Coriolis force depends upon the velocity, the random velocity distributions as defined by Kolmogorov depend not only upon the coordinates but also upon the initial velocity distributions. Therefore, the problem of defining simple and integrable spectral distributions of the Reynolds stresses that are linked to the kinetic energy spectrum is rendered completely inhomogeneous and therefore time dependent, breaking both of Kolmogorov's hypotheses that permit the construction of structure functions.

Nevertheless, there are parameter regimes in which approximate turbulence models of rotating thermal convection can be constructed. Depending upon that regime, the spectral scaling of the Reynolds stresses has been found to vary as $\widehat{v^2} \propto k^{\eta(k)}$, where $\eta(k)$ can range from being a constant \citep[e.g.][]{mininni10} to a function of wavenumber to account for various integral scales such as the Bolgiano scale \citep[e.g.][]{pharasi14}.
Thus, for simplicity and as a first approximation, the turbulence model presented in \cite{Samadietal2001}, \cite{Samadietal2010} and \citet{belkacem08} is adopted here, that is the Kolmogorov one, with the testing of more sophisticated rotating convection models being reserved for future work.

Therefore, we assume that the spatial structure of the convection is taken to be that of isotropic, homogeneous, and incompressible turbulence, which has a power law in the spatial transform of the kinetic energy for scales far from the integral driving scale and away from the diffusive microscale (Eq. \ref{turbulentspectrum}).  These assumptions that the motions are homogeneous and isotropic, have meaning only in a statistical sense.  That is to say that the turbulence model is built using random variables whose correlations are constructed such that they obey the expectation value of the energy equation applied to them \citep[we refer the reader to Eq. 14 in][]{Kolmogorov1941}. To paraphrase Kolmogorov and Taylor, this turbulence model can be considered in the following fashion: for very large Reynolds numbers, the first order perturbations on the average flow field consist of disorderly displacements of separate fluid volumes, one with respect to another, of a length scale $\ell_0$ (where $\ell_0$ is the Prandtl's mixing path length); these first-order perturbations are themselves of such a high amplitude that they are in turn unsteady and on these structures are superposed fluctuations of the second order with a mixing length $\ell_1<\ell_0$ and relative velocities $v_1<v_0$; such a process of successive refinement of turbulent perturbations may be carried out until for some sufficiently high order $n$ the Reynolds number $R_n = (l_n v_n)/\nu$, where $\nu$ is the kinematic viscosity, becomes so small that the effect of viscosity on the perturbations of the order $n$ finally prevents the formation of further fine structure at the order $n+1$.  To put that more succinctly, the spatial structure of the flow is a large scale flow with superposed self-similar eddies that obey the scale-by-scale averaged dissipation equation \citep[see again Eq. 14][]{Kolmogorov1941}.  Yet such motions are only captured in an ergodic statistical sense in the turbulence model of Kolmogorov used here.  

Nevertheless, the precise and time-evolving spectrum of the waves will be coupled to the actual structure of the turbulent flows that excite them, which are modified by rotation and not fully captured in the Kolmogorov's picture.  However, our measurements are integrated in time and space (over the disk of the star), and therefore represent an averaging operation, which is somewhat akin to the averaging procedure used to construct the Kolmogorov structured turbulence and certainly averages over some of the timescales of the convective motions.  So, as mentioned above, this is a first approximation for the modelling of the turbulence exciting the waves and additional theoretical efforts are ongoing to better address the interactions of waves with rotating turbulent structures from small to large scales \citep[e.g.][Augustson \& Mathis, submitted]{MNT2014,AugustsonMathis2019}, but they are beyond the scope of this work.

Following \cite{Samadietal2001}, we introduce a velocity $u_{0}$ and the entropy scalar variance
${\widetilde s}$ defined as

\begin{equation}
\frac{3}{2}u_{0}^2=\frac{1}{2}\Phi\,w^2=\int_{0}^{\infty}{\rm d}k\,E\left(k\right)
\label{w}
\end{equation}

\noindent and
\begin{equation}
\frac{1}{2}{\widetilde s}^2=\frac{1}{2}\left(\frac{2\Phi c_{p}}{g\Lambda\delta}\right)^2\,w^4=\int_{0}^{\infty}{\rm d}k\,E_{s}\left(k\right),
\end{equation}

\noindent where $w$ is the estimate for the vertical convective velocity deduced
from the mixing length theory \citep[MLT,][]{bohm1958}, while $\Phi$ is a factor
introduced by \cite{Gough1977} to take the anisotropy of the convective
turbulence into account. Here, we follow \cite{Samadietal2001} and we fix $\Phi=2$, which is the value that corresponds to the Bohm \& Vintense's MLT. $c_p$ is the heat capacity at constant pressure, $g$ is
the gravity and $\delta=\left(\partial\ln\rho_{0}/\partial\ln T_{0}\right)_{P}$.
Finally, $\Lambda=\alpha_{c}H_{p}$ is the usual mixing length, where $H_p$ is
the pressure scale height while the mixing-length parameter $\alpha_c=1.5$ in the computed stellar models which are used here to study the amplitude of stochastically-excited waves.

In addition, we define $k_0$ as the wavenumber of the largest eddy in the
inertial range. Following \cite{Samadietal2001}, we relate it to the mixing
length by

\begin{equation}
k_{0}=\frac{2\pi}{{\widehat \Lambda}}=\frac{2\pi}{{\widehat\beta}\Lambda},
\label{k0}
\end{equation}
where ${\widehat\beta}$ is a free parameter. \cite{Samadietal2001} demonstrated that for $\Phi=2$, ${\widehat\beta}\approx1.9$ (we refer the reader to their Eq. 76). However, \cite{Samadietal2010} suggested  choosing the size of the convective core for early-type stars instead of ${\widehat\beta}\Lambda$ since this is the natural size of the possible largest eddy \citep[see e.g. the 3D spherical numerical simulations by][]{Browningetal2004,augustson2016,Edelmannetal2019}. We hereafter adopt this choice for the computation of the amplitude of the waves.

The Kolmogorov spectrum is then expressed following \cite{Samadietal2001};

\begin{equation}
E\left(K\right)=\frac{u_{0}^{2}}{k_{0}}K^{-5/3},
\label{turbulentspectrum}
\end{equation}

\noindent where $K=k/k_{0}$, while
\begin{equation}
E_{s}\left(k\right)=\frac{1}{3}{\widetilde s}^{2}\frac{E\left(k\right)}{u_{0}^2}.
\end{equation}

Finally, we have to specify the time correlation function ($\chi_k$). As shown
by \cite{Samadietal2003} for solar p modes and by \cite{Belkacemetal2009} for
solar g modes, the choice of this function has a strong impact on the mode
amplitudes. In the case of massive stars, this has been confirmed by
\cite{Belkacemetal2010} and \cite{Samadietal2010}, who showed that the choice of
a Lorentzian function is the most favorable. Therefore, we follow 
\cite{Belkacemetal2010} and we define

\begin{equation}
\chi_{k}\left(\omega\right)=\frac{1}{\pi\omega_{k}}\frac{1}{1+\left(\frac{\omega}{\omega_k}\right)^2}
\end{equation}

\noindent with the normalisation condition
\begin{equation}
\int_{-\infty}^{\infty}\chi_{k}\left(\omega\right){\rm d}\omega=1.
\end{equation}

\noindent We define the linewidth
\begin{equation}
\omega_{k}=\frac{k\,u_{k}}{\lambda},
\end{equation}
where $\lambda$ is a free parameter that accounts for the uncertainties in
the definition of $\omega_{k}$ \citep[see][]{Samadietal2001,Belkacemetal2010},
while $u_{k}$ is the velocity of the eddy with the wavenumber $k$, which was
related to the kinetic energy spectrum by \cite{Stein1967}:

\begin{equation}
u_{k}^{2}=\int_{k}^{2k}{\rm d}k\,E\left(k\right).
\label{uk}
\end{equation}
Since the lifetime of the largest eddies in the inertial range 
\begin{equation}
\tau_{k_0}=\frac{\lambda}{\left(k_0 u_{k_0}\right)}
\end{equation}
cannot be longer than the characteristic time at which the convective energy dissipates into the turbulent cascade, we have
\begin{equation}
\tau_{k_0}\le\frac{\Lambda}{w}.    
\end{equation}
Using the definitions of $k_0$ (eq. \ref{k0}), $u_k$ (eq. \ref{uk}), and $w$ (eq. \ref{w}), we get $\lambda\le2.7\sqrt{\Phi}\approx3.82$; we here choose $\lambda=1$.

\end{appendix}
\end{document}